\documentclass{aa}

\usepackage[T1]{fontenc}
\usepackage{pslatex}
\usepackage{amsmath}
\usepackage{amsfonts}
\usepackage{xcolor}
\usepackage{url}
\usepackage{bm}
\usepackage{graphicx}
\usepackage{amssymb}
\usepackage{soul}
\usepackage{booktabs}
\usepackage{array}
\usepackage[utf8]{inputenc}
\inputencoding{latin1}
\inputencoding{utf8}
\usepackage{appendix}

\def\lsim{ \lower .75ex\hbox{$\sim$} \llap{\raise .27ex \hbox{$<$}} }
\def\gsim{ \lower .75ex \hbox{$\sim$} \llap{\raise .27ex \hbox{$>$}} }

\newcommand{\fig}[1]{Fig.~\ref{fig:#1}}
\newcommand{\eq}[1]{Eq.~(\ref{eq:#1})}

\newcommand{\unit}[1]{\nobreak{\mathrm{\;#1}}} 

\newcommand{\bi}{\begin{itemize}}
\newcommand{\ei}{\end{itemize}}

\usepackage[colorlinks=true,linkcolor=blue,citecolor=blue]{hyperref}

\begin{document}

\title{Stochastic acceleration in extreme TeV BL Lacs through MCMC} 

\author{
A. Sciaccaluga\inst{1,2}
\and F. Tavecchio\inst{2}
\and M. Landoni\inst{2}
\and A. Costa\inst{2,3}
}

\institute{
Dipartimento di Fisica, Università degli Studi di Genova, Via Dodecaneso 33, I-16146 Genova, Italy\\
\email{alberto.sciaccaluga@inaf.it}
\and
INAF -- Osservatorio Astronomico di Brera, Via E. Bianchi 46, I-23807 Merate, Italy
\and 
DiSAT, Università dell’Insubria, Via Valleggio 11, I-22100 Como, Italy
}
\date{}

\voffset-0.4in



\abstract
{Extreme TeV BL Lacs are a class of blazars with unique spectral and temporal features that are not easily reproducible using standard one-zone models based on single shock acceleration. To account for their peculiar properties, we elaborated a two-step acceleration model in which a recollimation shock and the subsequent downstream turbulence energize non-thermal electrons. 
}{We applied the model to a sample of extreme TeV BL Lacs with well-characterized spectral energy distributions. Since we used several sources, we automatized the exploration of the parameter space. This allowed us to derive the parameter distributions and study the correlations among them.}{We numerically solved a system of two coupled nonlinear differential equations to obtain the non-thermal particles and turbulence spectra. We calculated the spectral energy distribution via the synchrotron self-Compton emission model. The automatization of the parameter space exploration is possible through a Markov chain Monte Carlo (MCMC) ensemble sampler, in our case \tt{emcee}.}{We derived well-defined posterior distributions for the parameters, showing that the model is well constrained by available data and demonstrating the suitability of our method. The cross-correlations among some of the physical parameters are not trivial.\ Therefore, we conclude that MCMC sampling is a key instrument for characterizing the complexity of our multiparameter phenomenological model.}{}


\maketitle
\boldsymbol{}
\section{Introduction}
Active galactic nuclei (AGNs) are the most powerful persistent sources in the Universe. The output of these objects is powered by the gravitational energy released by gas accreting onto a supermassive black hole residing in a galactic nucleus. Radio-loud AGNs are characterized by the presence of a relativistic jet produced in the supermassive black hole vicinity that can propagate up to hundreds of kiloparsecs in the most powerful sources. Radio-loud AGNs are further divided into several classes based on their observational features, which are strongly dependent on the observational angle \citep{urrypadovani95}. Blazars are radio-loud AGNs whose jet is pointing toward the observer, and therefore the non-thermal emission of the jet dominates the emission thanks to relativistic Doppler beaming \citep{romero17, blandford19}. 

The spectral energy distribution (SED) of blazars presents two broad humps, the first due to synchrotron emission by non-thermal electrons. The origin of the second hump, peaking in the gamma-ray band, is still disputed: it could be generated by the interaction of non-thermal electrons with the synchrotron photons or with photons filling the external environment (i.e., synchrotron self-Compton and external Compton models; e.g., \citealt{ghisellini98}) or by hadronic processes, such as proton synchrotron emission or photo-pion production (e.g., \citealt{bottcher13}). 

Blazars can be classified using the frequencies of the two peaks (e.g., \citealt{ghisellini17}): the first ranges from the infrared to the X-ray bands and the second from MeV to TeV energies. The most efficient accelerators are the so-called extremely high-frequency-peaked BL Lacs (EHBLs; \citealt{costamante01}), which, in turn, can be divided into two subclasses, extreme synchrotron BL Lacs (which present the first peak above $1 \unit{keV}$) and extreme TeV BL Lacs. In addition, the latter are characterized by the second peak surpassing $1 \unit{TeV}$ and a hard GeV spectrum ($\Gamma < 2$ with $F_\nu/\nu \propto \nu^{-\Gamma}$). Furthermore, extreme TeV BL Lacs, at odds with the general behavior of blazars, present a low temporal variability at high energies \citep{biteau20}. 
The SEDs of these sources are difficult to explain through a leptonic single-zone model with a single shock acceleration: the spectral features imply a large minimum Lorentz factor, a small magnetic field far from equipartition, and a power law index incompatible with the theory of diffusive shock acceleration ($p < 2$ with $dN/dE \propto E^{-p}$). Several solutions have been proposed, such as a Maxwellian-like electron distribution \citep{lefa11}, a beam of high-energy hadrons \citep{essey10}, internal absorption \citep{aharonian08}, emission from a large-scale jet \citep{bottcher08}, lepto-hadronic models \citep{cerruti15}, and multiple shock acceleration \citep{zech21}.

To explain the phenomenology of extreme TeV BL Lacs, we elaborated a double-step acceleration model based on a scenario involving a jet recollimated by the pressure of external material. If the magnetization of the jet is low, which is likely for extreme TeV BL Lacs, the downstream of the shock formed as a consequence of the recollimation becomes unstable and turbulent \citep{gourgouliatos18,costa23}. In this scheme, non-thermal particles are first accelerated by the shock and then further energized in the downstream by turbulence through resonant interaction. Comparing turbulence cascading and damping timescales, it is evident that the damping cannot be neglected \citep{tavecchio22}.  In \cite{sciaccaluga22} we presented a time-dependent one-zone model that includes damping by the accelerating electrons. We calculated the nonthermal electron and turbulence spectra and then derived the SED using the synchrotron self-Compton (SSC) model. We compared our model with  data of the prototypical extreme TeV BL Lacs 1ES 0229+200 and adjusted, via visual inspection, our model on the flux points.

In this paper we apply the model sketched above but including other EHBLs with well-sampled SEDs. Moreover, to automatize and parallelize the comparison between the model and data, we developed a procedure based on a Markov chain Monte Carlo (MCMC) sampler. This technique allowed us to explore the parameter space of the model. 

The paper is organized as follows: in Sect. 2 we describe the code updates, in Sect. 3 we explain how we use MCMC sampling in our framework, and Sect. 4 is dedicated to the discussion. 
Throughout the paper, the following cosmological  parameters are assumed: $H_0=70{\rm\;km\;s}^{-1}{\rm\; Mpc}^{-1}$, $\Omega_{\rm M}=0.3$, and $\Omega_{\Lambda}=0.7$.

\section{The model}

Supposing a proton-electron plasma, the non-thermal electrons are the solely responsible for the emission. As described above, we assumed a model in which particles are accelerated at a shock and then energized through stochastic acceleration in the downstream, where they also emit through synchrotron and SSC mechanism. In our model, the non-thermal population of electrons accelerated at the shock is treated as an injection term in a kinetic equation for the stochastic acceleration. 

The time evolution of non-thermal electrons and turbulence is described by a system of two coupled nonlinear Fokker-Planck equations \citep{eilek79, miller95, kakuwa16}:
\begin{equation}
\left \{
\begin{aligned}
    & \frac{\partial f}{\partial t} = \frac{1}{p^2} \frac{\partial}{\partial p} \left [ p^2 D_p \frac{\partial f}{\partial p} + p^2 \left ( \frac{\partial p}{\partial t} \right )_{\text{rad}} f \right ] + \frac{f}{t_\text{esc}} +I_f \\
    & \frac{\partial Z}{\partial t} = \frac{1}{k^2} \frac{\partial}{\partial k} \left ( k^2 D_k \frac{\partial Z}{\partial k}  \right ) + \frac{Z}{t_\text{dam}} + \frac{I_W}{k^2}   
\end{aligned} \,.
\label{eq:system}
\right .
\end{equation}
Here $f(p,t)$ is the momentum distribution of non-thermal electrons, and $Z(k,t) = W/k^2$, where $W(k,t)$ is the wavenumber energy spectrum of turbulence, that is, the energy density (including both the kinetic and the magnetic component) of waves with wavenumber between $k$ and $k + dk$. \eq{system} includes all the physical processes related to non-thermal electrons and turbulence: particle resonant acceleration, cooling, escape, and injection together with turbulence cascading, damping, and injection. The two coupled equations are solved numerically using standard schemes (details in Appendix C). We used $20$ points per decade for the momentum and wavenumber grid, $50$ time steps, and $10$ points per decade for the frequency grid.

The momentum diffusion coefficient of the electrons is obtained from the quasi-linear theory of particle-wave interaction, supposing only parallel and antiparallel propagating magnetohydrodynamics  waves for a further simplification (e.g., \citealt{kakuwa16}):
\begin{equation}
    D_p = \frac{p^2 \,v_a^2}{U_B\, r_g^2\, c} \int_{k_\text{res}} \frac{W_B}{k}\,dk,
    \label{eq:Dp}
\end{equation}
where $v_a$ is the Alfvén velocity, $U_B = B^2/8\pi$ the total (ordered plus turbulent) energy density of the magnetic field, $m_e$ the electron mass, $c$ the light speed, $r_g = \gamma m_e c / eB$ the Larmor radius, $W_B \approx W/2$ the magnetic field component of the turbulence energy spectrum, and $k_\text{res} = 2\pi/r_g$ is the resonant wavenumber.

In addition to synchrotron cooling, this time we included Compton cooling for electrons, which increased the execution time but, after some optimizations, we were able to reach speeds (approximately a few seconds) comparable with other leptonic codes (e.g., \citealt{stathopoulos23}). For the synchrotron and Compton cooling, we used standard formulae. 

In our scenario, the escape depends on the turbulence. When the turbulence is strong, the electrons diffuse in the acceleration region, while if damping is important, the escape time is simply equal to the geometrical escape time. Therefore, the energy-dependent escape timescale can then be written as
\begin{equation}
    t_\text{esc} = \frac{R}{c} + \frac{R^2}{\kappa_\parallel},
\end{equation}
where $\kappa_\parallel = c r_g /9 \zeta(k_\text{res})$ is the spatial diffusion coefficient along the total magnetic field, while $\zeta(k) = kW_B/U_B$ is the relative amplitude of the turbulent magnetic field energy density for a given $k$.

The injection term describes the particles accelerated at the recollimation shock and advected downstream, supposing a strong shock:
\begin{equation}
    I_n = I_{n,0} \,\gamma^{-2}\, e^{-\frac{\gamma}{\gamma_\text{cut}}} \quad  \text{with} \quad  10^3 <\gamma<10^7,
\end{equation}
where $I_{n}$ is the injected electron number density per unit of time, which can be converted to the injection in the momentum space $I_n = 4\pi p^2 m_e c I_f$. The range of injection is determined from recent simulations of diffusive shock acceleration \citep{zech21}. The injection is normalized to the injected non-thermal electron power, $P_n$:
\begin{equation}
     P_n = V \int \gamma \, m_e\, c^2\, I_n \,d\gamma,
\end{equation}
where $V = 10\pi R^3$ is the emission region volume, modeled as a cylinder with radius $R$ and length $10\,R$. This estimate of the length for the emission volume, related to the region where the instability develops and triggers turbulence in the plasma, is roughly based on the simulations shown in \cite{matsumoto21}. We expect that within such a distance, both the magnetic field decay and the adiabatic losses effectively quench the emission  \citep{tavecchio22}. 

Regarding the turbulence equation, we used the Kolmogorov phenomenology, for which the effective diffusion coefficient is\begin{equation}
    D_k = \frac{1}{2}\,k^3\,v_a\,\sqrt{\frac{k\,W}{U_B}}.
\end{equation}
Without strong damping, with this diffusion coefficient and constant injection, $W(k)$ would reach a standard Kolmogorov spectrum, $W(k)\propto k^{-5/3}$ \citep{zhou90}.

The damping time is obtained by imposing energy conservation, that is, the energy used to accelerate the electrons is subtracted from the turbulence:
\begin{equation}
   t_\text{dam} = \left | \frac{4\pi e^2 v_a^2 }{m_e c^3 k} \int_{\gamma>\gamma_\text{res}} \gamma^2 \frac{\partial}{\partial \gamma} \left ( \frac{n_e}{\gamma^2} \right ) d\gamma \right |^{-1}.
\end{equation}
Here $n_e = 4\pi m_e c p^2 f$ is the electron energy spectrum and $\gamma_\text{res}$ is the resonant Lorentz factor, defined by $k = 2 \pi/r_g(\gamma_\text{res})$.

Finally, the turbulence is injected at a length $L$ equal to one-tenth of the emitting region radius, from which it cascades to shorter scales:
\begin{equation}
    I_W = I_{W,0}\, \delta (k-k_0).
\end{equation}
Here $\delta$ is the Dirac function and $k_0 = 2\pi/L$ is the injection wavenumber. The injection is normalized to the injected turbulence power, $P_w$:
\begin{equation}
     P_w = V \int I_W \,dk.
\end{equation}
More information on the different terms can be found in \cite{sciaccaluga22}.

\section{Markov chain Monte Carlo (MCMC)}

In the previous exploratory paper (\citealt{sciaccaluga22}) the model was adjusted on data by eye, a standard procedure in literature. However, this method presents several shortcomings, such as confirmation bias and repetition for each source. Therefore, we decided to move to MCMC sampling. This technique has several advantages: it automatizes and parallelizes the exploration of the parameter space, it can calculate the distribution and the cross-correlation of the model parameters and it permits the imposition of non-diagonal priors (i.e., conditions that depend on multiple parameters).

To explore the parameter space, we used the Python library \texttt{emcee} as the MCMC ensemble sampler \citep{emcee}. It implements several moves; we tested the default strech move \citep{goodman10} and the differential-independence mixture ensemble move \citep{boehl22}: with an equal number of steps, the latter presents a lower autocorrelation time, and we therefore  decided to adopt it. We assumed a Gaussian likelihood,
\begin{equation}
    \ln \mathcal{L} = \frac{1}{2} \sum_i \frac{ \left [(\nu\,F)_{i,m}-(\nu\,F)_i \right ]^2}{\sigma_i^2},
\end{equation}
where $(\nu\,F)_i$ and $\sigma_i$ are respectively the i-th flux point and its uncertainty, while $(\nu\,F)_{i,m}$ is the model flux at the i-th frequency. The model flux is calculated using standard synchrotron and SSC emissivities, after integrating \eq{system} until $t_{\rm max}=10R/c$ (see \citealt{sciaccaluga22} for the details). In literature, it is a standard procedure to add a term in the likelihood to account for the scatter due to the non-simultaneity of flux points \citep{hogg10, stathopoulos23}, but extreme TeV BL Lacs are highly stable. Tests that included this nuisance parameter resulted in extremely low values; therefore, we neglected it. 

\begin{table}[t]
\caption{Recap parameter table, indicating their symbols, units, ranges, and prior distributions}
\resizebox{\columnwidth}{!}{
\begin{tabular}{lllcc}
\hline
Parameter                 & Symbol & Units & Range                & Type of prior \\ \hline
Emitting region radius    & $R$    & cm    & $[10^{14}, 10^{20}]$ & Log-flat      \\
Alfvén velocity           & $v_a$  & cm/s  & $[10^8,10^{10}]$     & Log-flat    \\
Total magnetic field       & $B$    & G     & $[10^{-5},10^0]$     & Log-flat      \\
Electrons injected power  & $P_n$  & erg/s & $[10^{35},10^{44}]$  & Log-flat      \\
Turbulence injected power & $P_w$  & erg/s & $[10^{38},10^{42}]$  & Log-flat \\ \hline 
\end{tabular}
}
\label{tab:priors}
\end{table}

\subsection{Fixing priors}

Our model presents several parameters: the radius of the emission region, $R$, the Alfvén velocity, $v_a$, the total magnetic field, $B$, the non-thermal electron and turbulence injection power in the jet frame, respectively $P_n$ and $P_w$, and the relativistic Doppler factor, $\delta$.

After some tests, we decided to fix the relativistic Doppler factor to $\delta = 20$. If left free, it tends to reach $\delta \sim 100$, a value that we consider implausible for a blazar. On the other hand, the relativistic Doppler factor cannot be too small, since it would result in small boosting and in a large emission region radius, again implausible for a blazar region (at sub-parsec scales). We were finally left with five free parameters.

As shown in Table \ref{tab:priors}, we applied log-flat priors to all parameters over broad ranges, allowing the walkers to explore a large part of the parameter space. Furthermore, we imposed two non-diagonal priors. Since the interaction between turbulence and non-thermal particles is described using quasi-linear theory,  we required the turbulent component to be small compared to the total field in the emission region. Moreover, we required the non-thermal component to be a minor part of the post-shock plasma (e.g., \citealt{sironi11}). Specifically, we put constraints on the ratio of the energy density of the magnetic turbulence and of the total field, $U_B$, and on the ratio of the number density of the non-thermal electrons, $N_e$, and of the thermal plasma, $N_p$, calculated from the Alfvén velocity and the total magnetic field:
\begin{equation}
    \begin{aligned}
        & \frac{\delta B^2}{B^2} = \frac{\int W_B(k)\, dk}{U_B} < 0.1 \\
        & \frac{N_e}{N_p} < 0.1.
    \end{aligned}
\end{equation}
No strong limits were imposed on the magnetization $\sigma = B^2/8\pi m_p N_p$, where $m_p$ is the proton mass. It was used as a consistency check since we expect a low value, necessary for the development of the instabilities in the downstream region.

\section{Results}
We applied the procedure described above to four extreme TeV BL Lacs taken from \cite{costamante18}: 1ES 0229+200, 1ES 0347-121, RGB J0710+591, and 1ES 1101-232.  All these EHBLs have a well-characterized SED with a synchrotron peak well tracked by \textit{Swift} and \textit{NuSTAR} and a high-energy peak covered by \textit{Fermi} and Cherenkov telescopes. From the sample of \cite{costamante18}, comprising a total of six sources, we did not consider two EHBLs, namely 1ES 0414+009 and 1ES 1218+304. For the former object, the slope of the \textit{Fermi} and the TeV spectra are apparently in disagreement, making the SED unsuitable for the automated fit approach. For 1ES 1218+304 the situation is rather complex: in the optical-UV region, the relative contribution of the host galaxy and the jet is not easy to disentangle. Moreover, the \textit{Swift} data do not clearly trace the shape of the low-energy part of the synchrotron peak. For these reasons, we preferred to exclude also this complex source from our study.

In the X-ray band, the model uncertainty is smaller than in other bands since \textit{Swift} and \textit{NuSTAR} measurements are more accurate than \textit{Fermi} and Cherenkov arrays. It is also worth noticing that the \textit{Fermi} spectrum is well reproduced for each source, thanks to the softening due to the turbulence damping \citep{sciaccaluga22}. Finally, very high-energy data are compatible within $1\sigma$ uncertainty, except for the most energetic flux points of 1ES0229+200 and 1ES 0347-121, which are still compatible at $<2\sigma$. In no cases did we consider the optical-UV data in the fit, since in these bands the emission is dominated by the host galaxy.

\begin{figure}[h]
    \centering
    \includegraphics[width = \columnwidth]{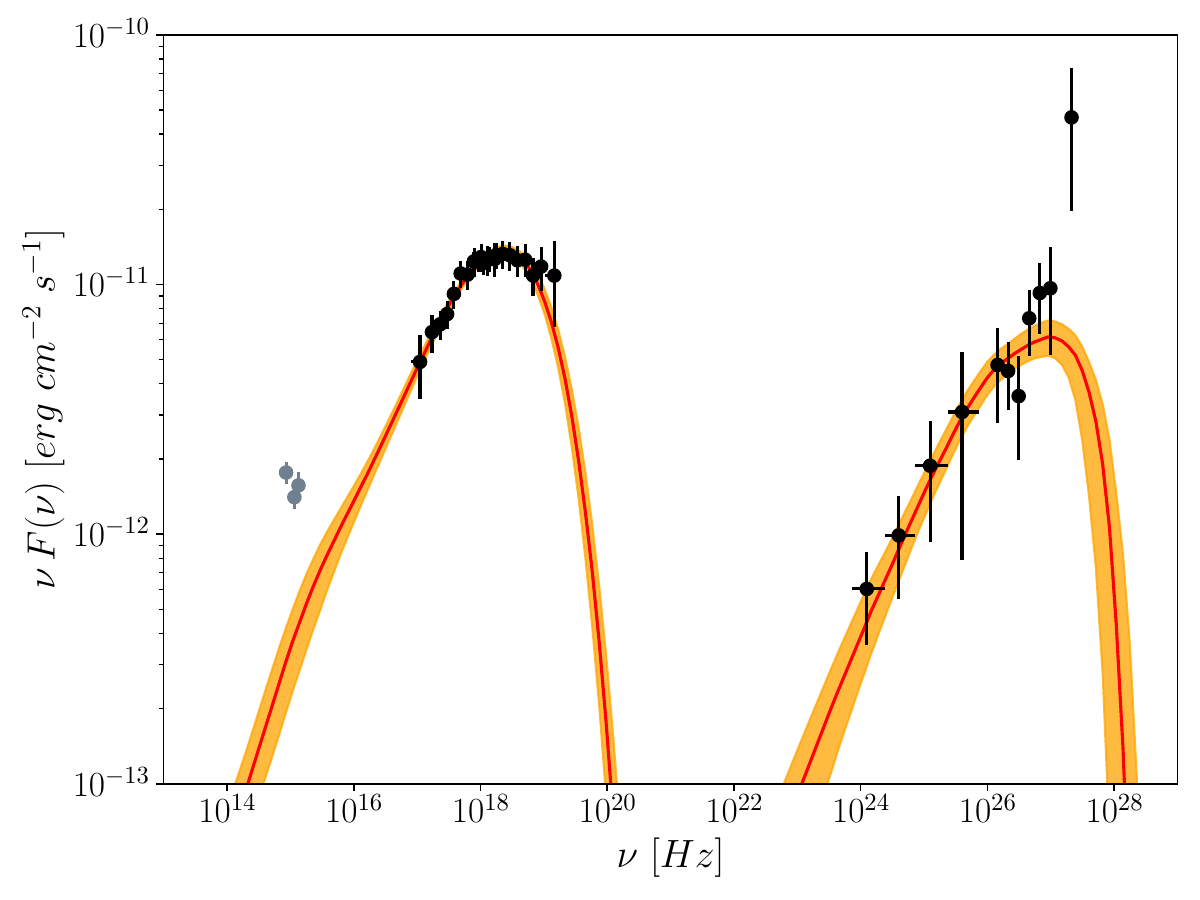}
    \caption{Flux points with their errors (black) of 1ES 0229+200 with $1\sigma$ credible intervals (orange) and the median (red) obtained from the model posterior. Gray points correspond to \textit{Swift}/UVOT data.}
    \label{fig:flux0229}
\end{figure}

\begin{figure}[h]
    \centering
    \includegraphics[width = \columnwidth]{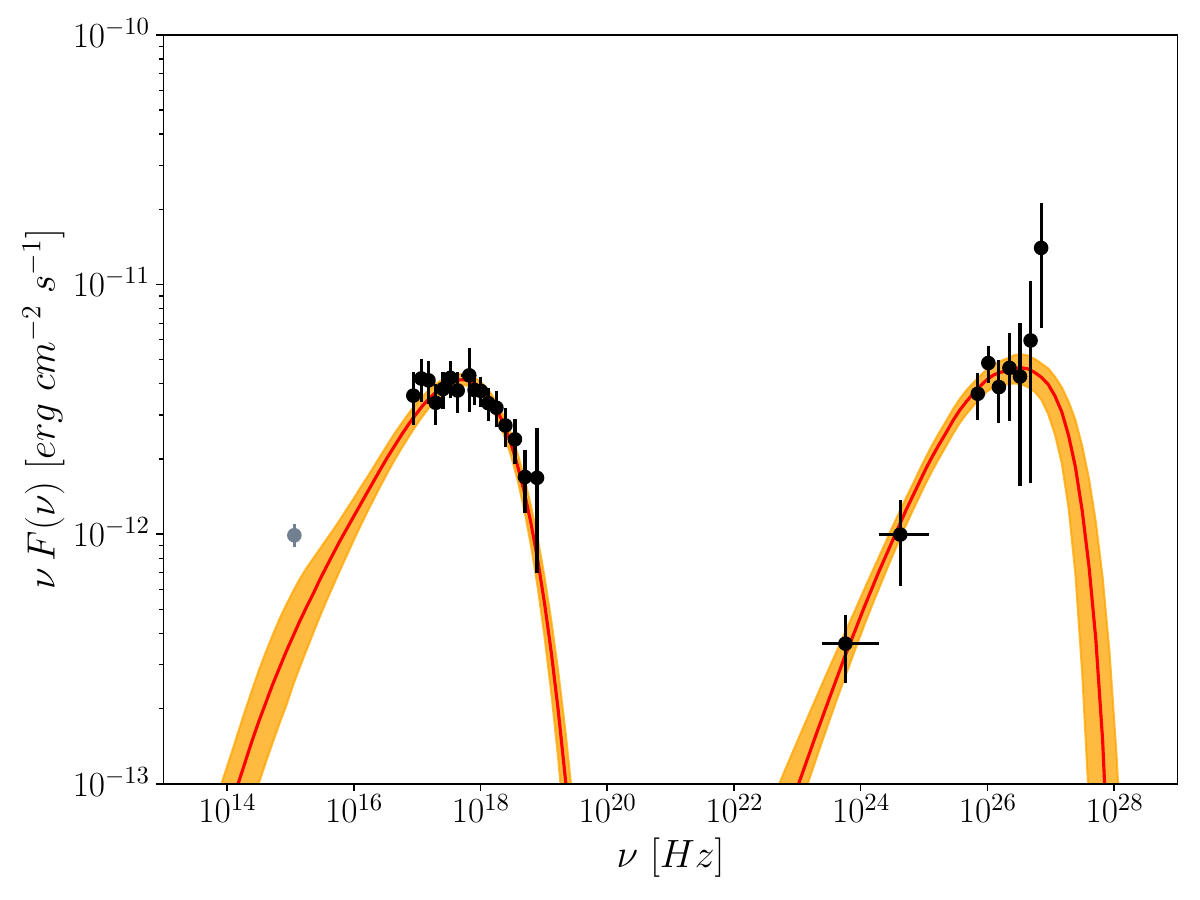}
    \caption{Flux points with their errors (black) of 1ES 0347-121 with $1\sigma$ credible intervals (orange) and the median (red) obtained from the model posterior. Gray points correspond to \textit{Swift}/UVOT data.}
    \label{fig:flux0347}
\end{figure}

\begin{figure}[h]
    \centering
    \includegraphics[width = \columnwidth]{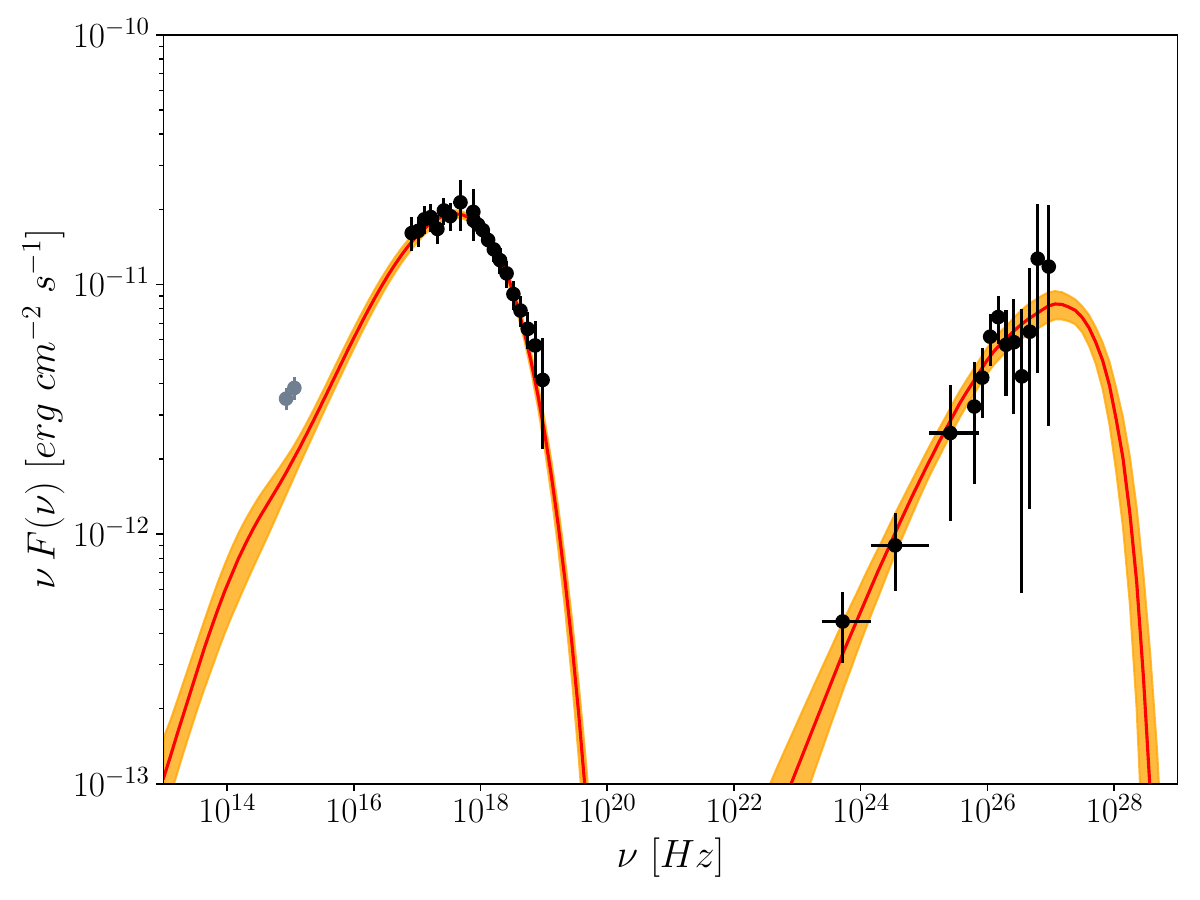}
    \caption{Flux points with their errors (black) of 1ES 1101-232 with $1\sigma$ credible intervals (orange) and the median (red) obtained from the model posterior. Gray points correspond to \textit{Swift}/UVOT data.}
    \label{fig:flux1101}
\end{figure}

\begin{figure}[h]
    \centering
    \includegraphics[width = \columnwidth]{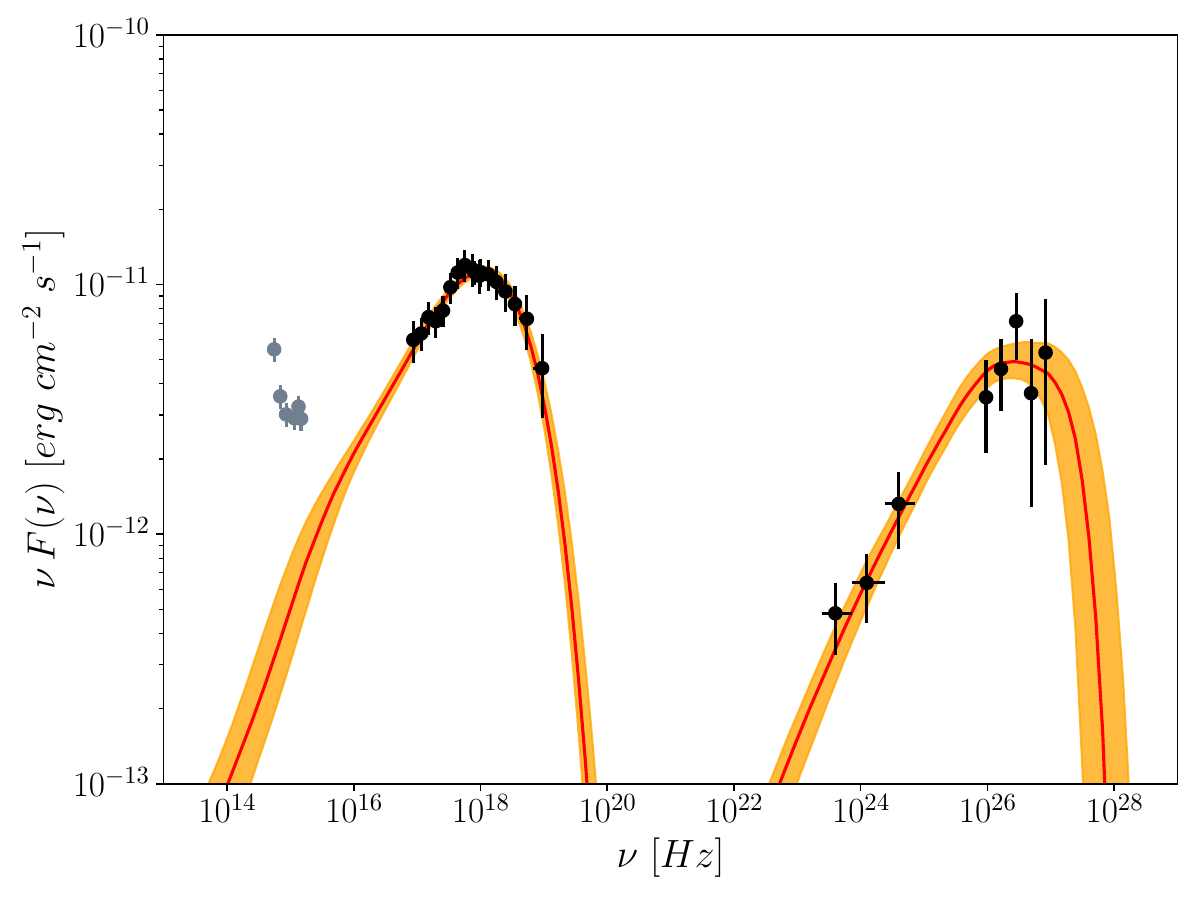}
    \caption{Flux points with their errors (black) of RGB J0710+591 with $1\sigma$ credible intervals (orange) and the median (red) obtained from the model posterior. Gray points correspond to \textit{Swift}/UVOT data.}
    \label{fig:flux0710}
\end{figure}

For each source, we initialized $30$ walkers, each moving for $10000$ steps and with $\sim 1000$ burn-in steps. In Figs. \ref{fig:flux0229}, \ref{fig:flux0347}, \ref{fig:flux1101}, and \ref{fig:flux0710} the measured flux data points are shown together with the median and the $1\sigma$ credible interval of the calculated SEDs, obtained drawing $\sim 1000$ random samples from the posterior. The corresponding corner plots, showing the distribution of the derived model parameters, are shown in Appendix A.

Examining the corner plots (see Figs. \ref{fig:CP0229}, \ref{fig:CP0347}, \ref{fig:CP1101}, and \ref{fig:CP0710}), the distributions in logarithmic space appear narrow (evidence of a good reconstruction of the physical parameters), but strongly not-gaussian. They peak in the same range of parameters, with slightly broader uncertainties for RGB J0710+591 and with 1ES 1101-232 as the only exception. The emitting region radius ($R\sim 10^{16 \div 17}\unit{cm}$) and the magnetic field ($B\sim 10^{-3 \div -2}\unit{G}$) are compatible with the values obtained by phenomenological models that leave the acceleration process unspecified (e.g., \citealt{costamante18}). The Alfvén velocity ($v_a\sim 10^{9}\unit{cm/s}$) aligns with values obtained in literature (e.g., \citealt{kakuwa16, tavecchio22}). Finally, the injected electron power ($P_n\sim 10^{40\div 41}\unit{erg/s}$) is consistent with values utilized with other BL Lacs (e.g., \citealt{ghisellini10}).  

For  1ES 1101-232 the fit provides a radius much larger than the value derived for the other sources ($R \sim 1 \unit{pc}$), together with a smaller magnetic field and higher electron and turbulence powers. These differences are probably connected to the steeply decreasing X-ray spectrum (i.e., a relatively low synchrotron peak frequency), a feature absent in the other sources. A sub-parsec-sized emission region can be obtained by applying a larger relativistic Doppler factor ($\delta \sim 40$). The results are shown in Appendix B. 

Quite interestingly, the corner plots reveal strong cross-correlations among some of the parameters, which are hard to discover through the standard approach based on the visual comparison of the model and data. For example, there is a strong anticorrelation between the radius of the emitting region and the magnetic field for each source. This is explained by the fact that the magnetic field determines the magnetic field energy density $U_B$, while, for a fixed synchrotron luminosity, the radius regulates the radiation energy density, $U_{\rm rad}$. Since the ratio $U_B/U_{\rm rad}$ is fixed by the ratio of the synchrotron and SSC peak luminosities (e.g., \citealt{tavecchio98}), this fixes the product $BR$. While this correlation is strong for each source and can be directly interpreted, some other correlations are weaker and much more difficult to explain in simple terms, because they are associated with observational features that involve several parameters.

\begin{figure}[h]
    \centering
    \includegraphics[width = \columnwidth]{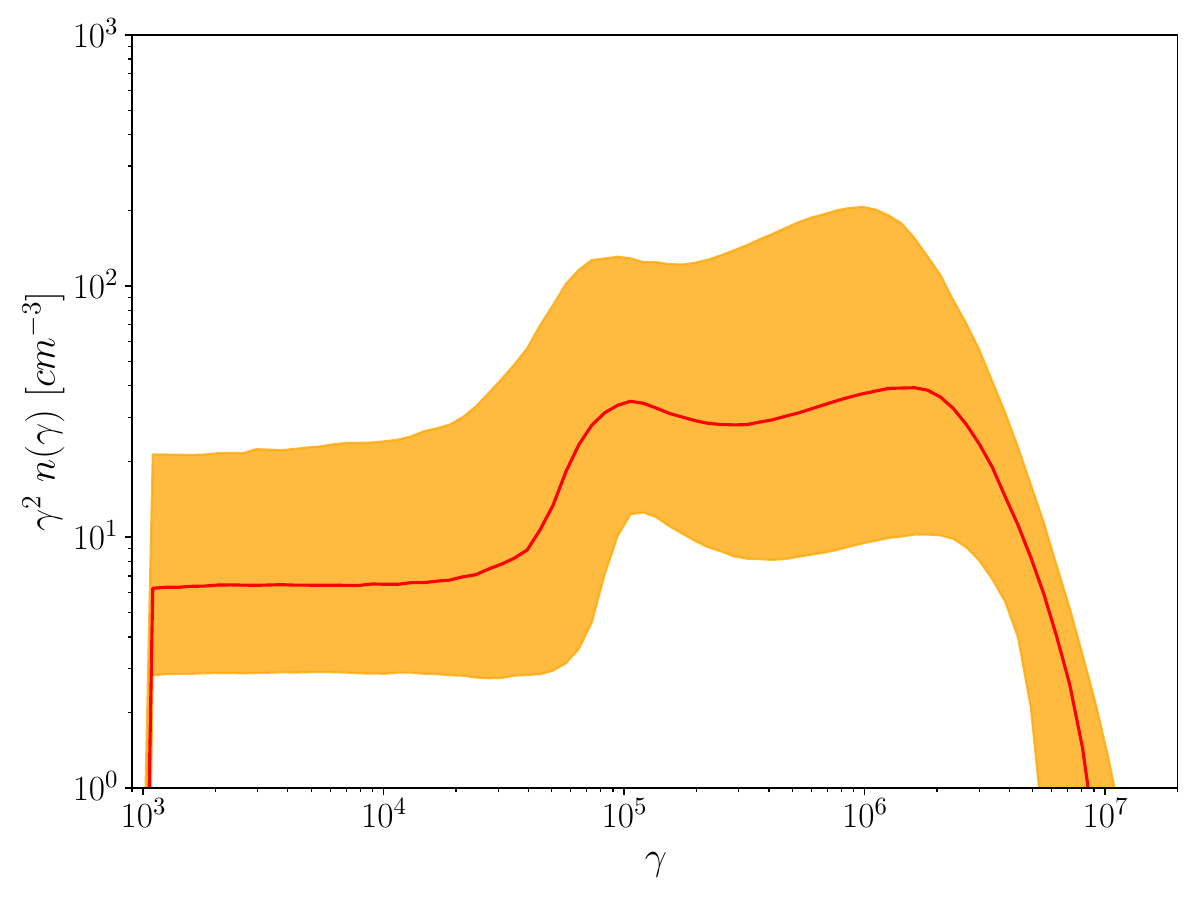}
    \caption{Final electron energy distribution for 1ES 0229+200 with $1\sigma$ credible intervals (orange) and the median (red) obtained from the model posterior.}
    \label{fig:electron0229}
\end{figure}

\begin{figure}[h]
    \centering
    \includegraphics[width = \columnwidth]{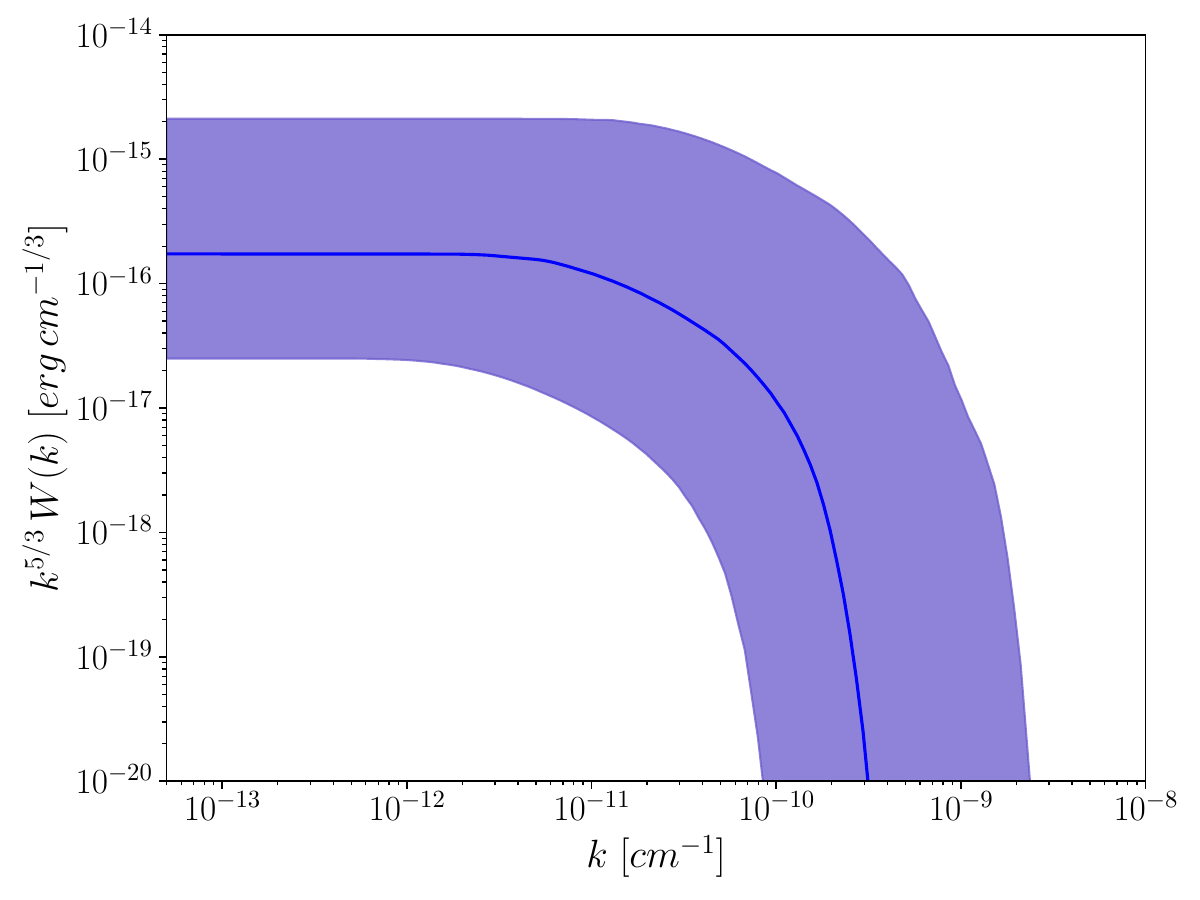}
    \caption{Final spectrum of the turbulence for 1ES 0229+200 with $1\sigma$ credible intervals (violet) and the median (blue) obtained from the model posterior.}
    \label{fig:turbulence0229}
\end{figure}

\begin{figure}[h]
    \centering
    \includegraphics[width = \columnwidth]{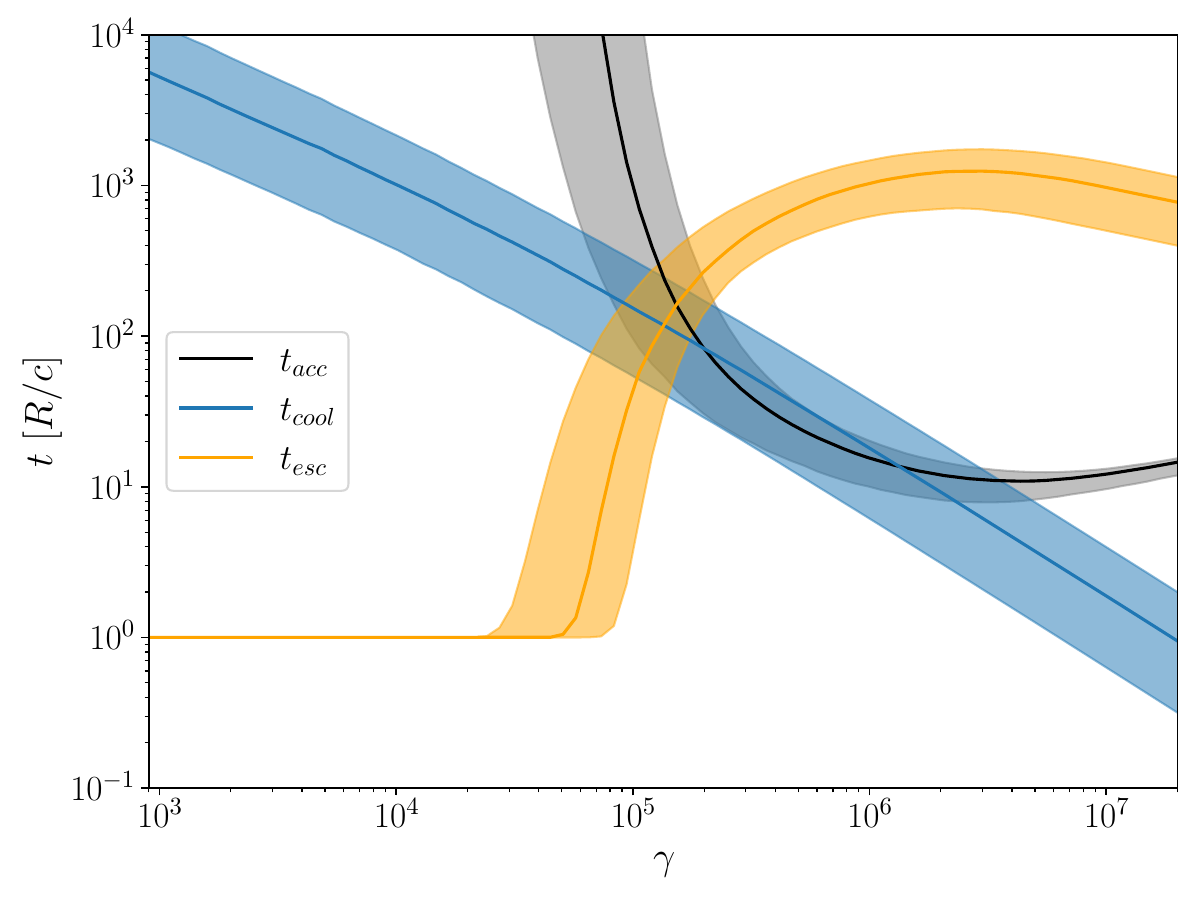}
    \caption{Final electron timescales (acceleration, cooling, escape) for 1ES 0229+200 with $1\sigma$ credible intervals and the median obtained from the model posterior.}
    \label{fig:electron_times_0229}
\end{figure}

\begin{figure}[h]
    \centering
    \includegraphics[width = \columnwidth]{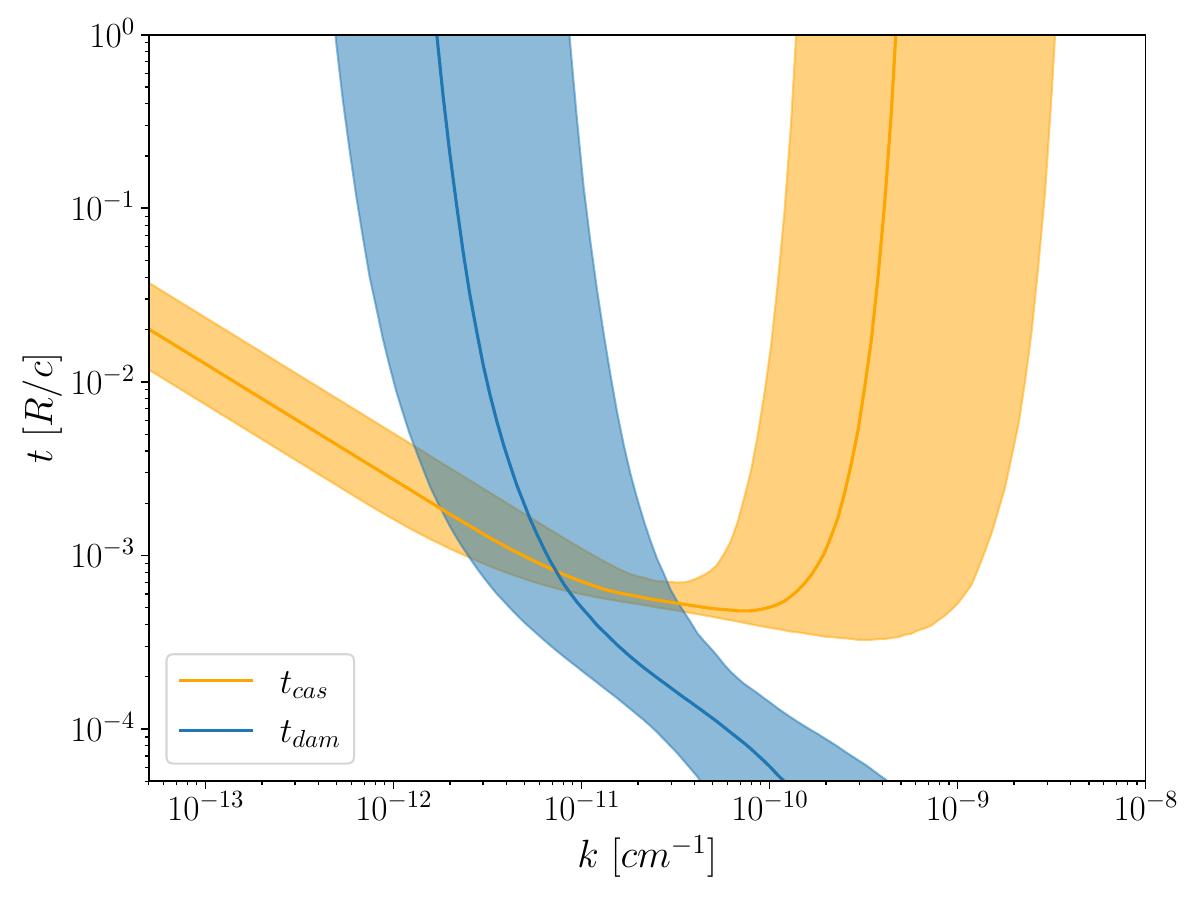}
    \caption{Final turbulence timescales (cascading and damping) for 1ES 0229+200 with $1\sigma$ credible intervals and the median obtained from the model posterior.}
    \label{fig:turbulence_times_0229}
\end{figure}

We calculated the final electron energy distribution, $n(\gamma)$, and the final turbulence spectrum, $W(k)$, (medians and credible intervals), drafted from the posterior. As a representative case, we display in \fig{electron0229} and \fig{turbulence0229} the results for 1ES 0229+200.  Finally, we report the median and the credible interval of the final timescales associated with electrons (i.e., acceleration, cooling, and escape time) and turbulence (cascading and damping times) for 1ES 0229+200, respectively in \fig{electron_times_0229} and \fig{turbulence_times_0229}.

As discussed in \cite{sciaccaluga22}, the combined effects of acceleration and radiative cooling result in a final electron energy distribution displaying a well-defined peak that, for EHBLs, is generally located at Lorentz factors $\gamma_p\approx 10^6$, where $t_{\rm acc}\sim t_{\rm cool}$ (see Fig. \ref{fig:electron_times_0229}). The role of turbulence acceleration is relevant only at high energies,  $10^5 \lesssim \gamma \lesssim 10^6$, where it piles up the injected electrons. At a lower energies ($\gamma\lesssim 10^5$), particles are inefficiently accelerated and effectively escape from the acceleration and emission region. Since at these energies electrons are unaffected by acceleration or cooling ($t_{\rm esc}\ll t_{\rm cool}$ and $t_{\rm esc}\ll t_{\rm acc}$), the distribution tends to be similar to that  injected by the shock (i.e., $\gamma^{-2}$). 

For turbulence, damping is effective only at large $k$, where the damping time is less than the cascading time (Figs. \ref{fig:turbulence0229} and \ref{fig:turbulence_times_0229}). At these wavenumbers, the damping is a consequence of the large number of resonating low-energy electrons. High-energy electrons ($\gamma\gtrsim 10^5$), responsible for the X-ray emission, resonate with low $k$ modes ($k<10^{-11}$ cm$^{-1}$), where, instead, damping is not efficient due to their small number. Notably, the very inefficient acceleration of the electrons at low Lorentz factors at late times (clearly visible in Fig. \ref{fig:electron_times_0229}) is related to the strong damping of the turbulence at large $k$.

The electron energy distribution shows a large uncertainty (about an order of magnitude) linked to the complex interplay between the electron density, the peak energy of the electron energy distribution, the magnetic field and the radius of the emitting region. The uncertainty on the normalization of the electron density, $n$, can be understood 
recalling that, as discussed above, the product $BR$ is fixed by the observed ratio of the synchrotron and SSC peak luminosity. This, for a fixed synchrotron luminosity, proportional to $L_S\propto n B^2 R^3$, implies that $n$ and $R$ are anticorrelated: $nR=$const. Therefore, a large (small) radius implies both a low (high) density and a low (high) magnetic field. In order to have a fixed synchrotron peak frequency, proportional to $\nu_s\propto \gamma _p^2 B$, the electron distribution must peak at high (low) Lorentz factor $\gamma_p$ (as visible in Fig. \ref{fig:electron0229}). The large uncertainty of the electron distribution is also reflected by the SSC spectrum, which, however, presents a  smaller variance. The same SSC spectrum can be obtained by using different electron spectra and target photon fields, therefore reducing the SSC spectrum uncertainty at the expense of less constrained optical and radio spectra.   

The turbulence energy spectrum displays a large uncertainty as well. In fact, the same acceleration efficiency can be obtained with a high magnetic field (or Alfvén velocity) and low turbulence injection and vice versa (see Eq. \ref{eq:Dp} and the corner plot). The much larger uncertainty in the region of $k$ where the spectrum is affected by damping reflects the strong interaction with the electrons.

\begin{figure}[h]
    \centering
    \includegraphics[width = \columnwidth]{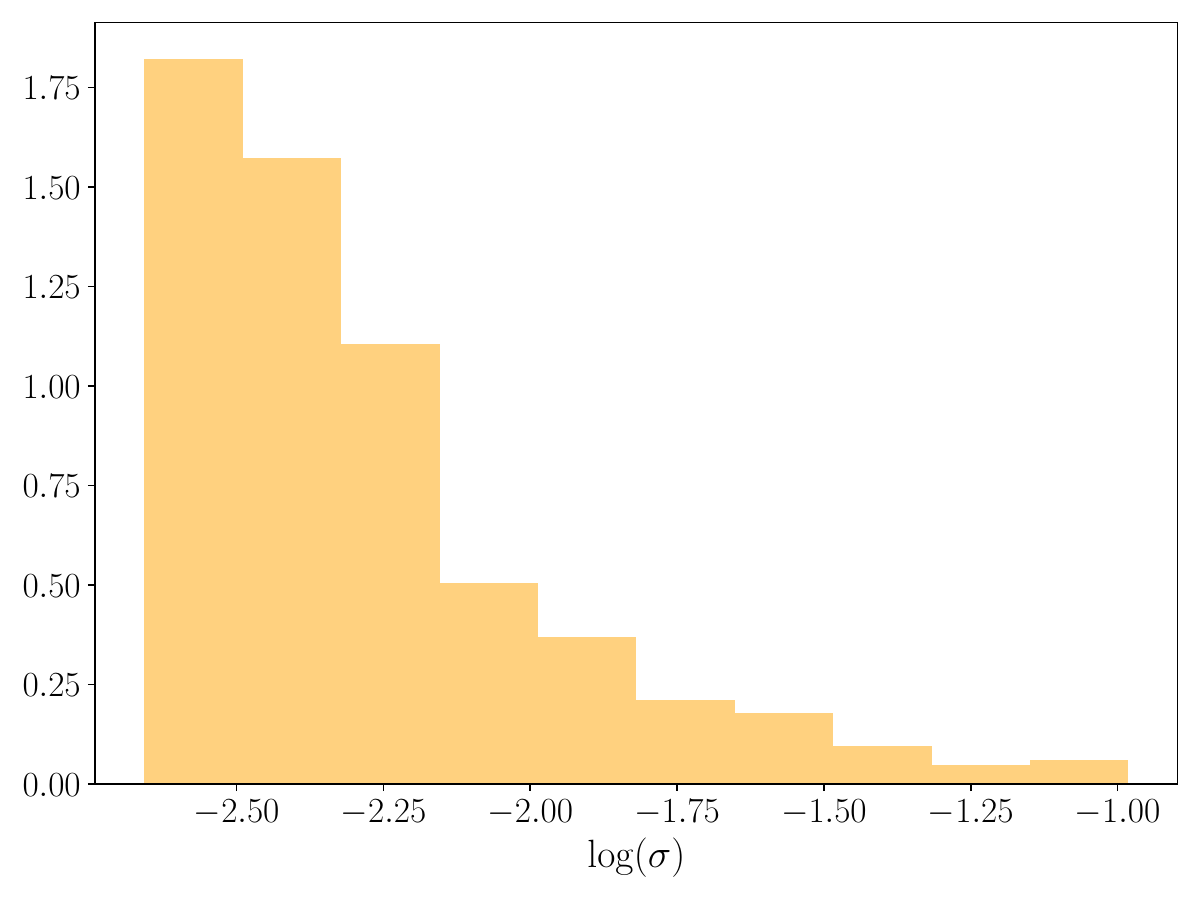}
    \caption{Distribution of the magnetization of 1ES0229+200 obtained from the model posterior.}
    \label{fig:magn0229}
\end{figure}

Starting from the Alfvén velocity, we calculated the magnetization distributions to confirm the consistency of our model. Using 1ES0229+200 as a representative case, we obtain a low magnetization $\sigma<10^{-2}$ (see \fig{magn0229}). Such low values are compatible with estimates from simple leptonic scenarios (e.g., \citealt{zech21}) and are below the threshold derived for the development of instabilities in the downstream region of the recollimation shock \citep{gourgouliatos18}.

\section{Discussion}

We have presented the modeling of a sample of EHBLs with well-reconstructed SEDs using a double acceleration (shock+stochastic) model \citep{tavecchio22}. By means of the MCMC technique, we have explored the full parameter space.
The MCMC sampling permitted us to approach the modeling without confirmation biases and showed us that the different parameters are correlated in a nontrivial way. This confirms that the standard simple visual comparison between the data and one realization of the model is strongly limited and that sampling is necessary to completely characterize the model complexity. Indeed, while the most probable values of the derived model parameters are in line with those derived with the standard approach, in the present work we were able to infer the related uncertainties and the correlations among them. We confirm for all sources that the interplay between the turbulence and the accelerating electrons leads to a strong damping of the turbulent energy at large wave vectors. From the point of view of the energetic budget, the scenario requires moderate powers in the form of injected turbulence and electrons, which is completely compatible with the jet power derived by means of standard one-zone models \citep{ghisellini10}.

In the original version of our scenario \citep{tavecchio22} we assumed that the stochastic acceleration occurs in the turbulent region after a recollimation shock. However, constraining the size of the emission region to sub-parsec scales results in a relatively large relativistic factor ($\delta = 20-40$). Such values, although  modest compared to those derived with simple one-zone leptonic models (e.g., \citealt{costamante18}), seem to challenge the scenario in which turbulence energizing the pre-accelerated electrons is the result of global instabilities excited by recollimation, since these instabilities are expected to strongly affect the flow, eventually disrupting it. For instance, in the simulations of \cite{matsumoto21} and \cite{costa23}, the turbulent post-shock flow is strongly decelerated and the average bulk Lorentz factor is only a few, implying a moderate beaming of the radiation. Another factor that potentially impacts our scenario was revealed in recent results from the Imaging X-ray Polarimetry Explorer (IXPE). Observations of 1ES 0229+200 \citep{ixpe23} show that the X-ray polarization is rather high, $\sim 18\%$, and strongly chromatic (i.e., X-ray polarization is higher than optical polarization). In our previous work, we suggested that in the model it is natural to expect a low polarization in all bands, but more accurate estimates are necessary in light of these detailed measurements. Following the phenomenological approach of \cite{marscher22}, the mean polarization degree of a turbulent emitting region is equal to
\begin{equation}
    \langle P \rangle \approx 0.75 [f_{\text{ord}}^2 + (1-f_{\text{ord}})^2 N_{\text{cell}}^{-1}]^\frac{1}{2},
\end{equation}
where $f_\text{ord} = B_\text{ord}/B$ (with $B_\text{ord}$ the ordered component of the magnetic field) and $N_\text{cell}$ is the number of turbulent cells. Fixing $f_\text{ord}$, the minimum is reached when $N_\text{cell} \to \infty$; therefore, the maximum value of $f_\text{ord}$ necessary to obtain the polarization degree observed by IXPE is given by $f_\text{ord,max} = \langle P \rangle / 0.75 \approx 0.24$, which implies $\delta B^2/B^2 \gsim 0.6$, a value beyond the scope of the quasi-linear theory at the base of our treatment of stochastic acceleration. 

In view of these difficulties, we propose a modification of the scenario that can potentially account for both the moderate relativistic Doppler factor and the chromatic polarization.
While in the original idea the post-shock turbulence is supposed to be related to the onset of global jet instabilities triggered by recollimation, an alternative is to assume the existence of inhomogeneities in the upstream flow. This, in particular if the plasma is characterized by a low magnetization, promotes the development of Richtmyer--Meshkov-like instability at the shock, with the resulting onset of turbulence in the downstream region. Importantly, dedicated magnetohydrodynamic  simulations show that close to the  shock front the evolving turbulent eddies are small, but their size grows moving away from the shock (e.g., \citealt{mizuno14}). Since the jet does not mix with the external medium, it can maintain a moderate relativistic Doppler factor, as required by our fits. Moreover, the structure of the turbulence can explain chromatism: low-energy electrons that emit in the optical band are accelerated by small eddies (i.e., eddies characterized by a large $k$), which are present both near the shock front and far from it since large eddies naturally cascade at shorter lengths. On the other hand, high-energy electrons emitting in the X-ray band have large Larmor radii and are associated exclusively with the eddies with large wavelengths (small $k$) far from the shock front. We therefore expect that the effective volume where X-ray emission occurs is much smaller than that associated with the optical band, implying an X-ray polarization higher than the optical one, in line with observations \citep{marscher22}. An improvement to our model could be the introduction of a time-dependent injection, to model the increase in the eddy size away from the shock front. Simulations are needed to understand the complete phenomenology of extreme TeV BL Lacs. Thus, our next step will be to use the Lagrangian particle framework to investigate these sources \citep{vaidya18, mukherjee21}. 

The next generation of gamma-ray facilities, such as CTA and ASTRI, will provide more constraining measurements that will further reduce the model uncertainty. In particular, our model predicts a strong cutoff above $10 \unit{TeV}$ that, if confirmed, will strongly limit the use of extreme TeV BL Lacs as beacons to test physics beyond the standard model (see, e.g., \citealt{galanti20}), such as axion-like particles and Lorentz invariance violation. 

The exploration of the parameter space for phenomenological models has recently improved thanks to neural networks. After being trained on a sample of several SEDs, neural networks are capable of computing spectra in milliseconds, instead of the few seconds required by leptonic codes. Together with MCMC or nested sampling, neural networks can be used to efficiently explore the parameter space \citep{begue23, tzavellas23}. We aim to implement this technique for our model as well. 

\begin{acknowledgements}

We thank A. Mignone and S. Kundu for useful discussions on the numerical schemes. We acknowledge financial support by a INAF Theory Grant 2022 (PI F. Tavecchio) and the PRIN 2022 (2022C9TNNX) project. This work has been funded by the EU - Next Generation EU.\end{acknowledgements}

\bibliographystyle{aa}
\bibliography{biblio}

\begin{thebibliography}{45}
\expandafter\ifx\csname natexlab\endcsname\relax\def\natexlab#1{#1}\fi

\bibitem[{{Aharonian} {et~al.}(2008){Aharonian}, {Khangulyan}, \& {Costamante}}]{aharonian08}
{Aharonian}, F.~A., {Khangulyan}, D., \& {Costamante}, L. 2008, \mnras, 387, 1206

\bibitem[{{B{\'e}gu{\'e}} {et~al.}(2023){B{\'e}gu{\'e}}, {Sahakyan}, {Dereli B{\'e}gu{\'e}}, {Giommi}, {Gasparyan}, {Khachatryan}, {Casotto}, \& {Pe'er}}]{begue23}
{B{\'e}gu{\'e}}, D., {Sahakyan}, N., {Dereli B{\'e}gu{\'e}}, H., {et~al.} 2023, arXiv e-prints, arXiv:2311.02979

\bibitem[{{Biteau} {et~al.}(2020){Biteau}, {Prandini}, {Costamante}, {Lemoine}, {Padovani}, {Pueschel}, {Resconi}, {Tavecchio}, {Taylor}, \& {Zech}}]{biteau20}
{Biteau}, J., {Prandini}, E., {Costamante}, L., {et~al.} 2020, Nature Astronomy, 4, 124

\bibitem[{{Blandford} {et~al.}(2019){Blandford}, {Meier}, \& {Readhead}}]{blandford19}
{Blandford}, R., {Meier}, D., \& {Readhead}, A. 2019, \araa, 57, 467

\bibitem[{Boehl(2022)}]{boehl22}
Boehl, G. 2022, Ensemble MCMC Sampling for Robust Bayesian Inference, Tech. rep.

\bibitem[{{B{\"o}ttcher} {et~al.}(2008){B{\"o}ttcher}, {Dermer}, \& {Finke}}]{bottcher08}
{B{\"o}ttcher}, M., {Dermer}, C.~D., \& {Finke}, J.~D. 2008, \apjl, 679, L9

\bibitem[{{B{\"o}ttcher} {et~al.}(2013){B{\"o}ttcher}, {Reimer}, {Sweeney}, \& {Prakash}}]{bottcher13}
{B{\"o}ttcher}, M., {Reimer}, A., {Sweeney}, K., \& {Prakash}, A. 2013, \apj, 768, 54

\bibitem[{{Cerruti} {et~al.}(2015){Cerruti}, {Zech}, {Boisson}, \& {Inoue}}]{cerruti15}
{Cerruti}, M., {Zech}, A., {Boisson}, C., \& {Inoue}, S. 2015, \mnras, 448, 910

\bibitem[{{Chang} \& {Cooper}(1970)}]{chang70}
{Chang}, J.~S. \& {Cooper}, G. 1970, Journal of Computational Physics, 6, 1

\bibitem[{{Costa} {et~al.}(2023){Costa}, {Bodo}, {Tavecchio}, {Rossi}, {Capetti}, {Massaglia}, {Sciaccaluga}, {Baldi}, \& {Giovannini}}]{costa23}
{Costa}, A., {Bodo}, G., {Tavecchio}, F., {et~al.} 2023, arXiv e-prints, arXiv:2312.08767

\bibitem[{{Costamante} {et~al.}(2018){Costamante}, {Bonnoli}, {Tavecchio}, {Ghisellini}, {Tagliaferri}, \& {Khangulyan}}]{costamante18}
{Costamante}, L., {Bonnoli}, G., {Tavecchio}, F., {et~al.} 2018, \mnras, 477, 4257

\bibitem[{{Costamante} {et~al.}(2001){Costamante}, {Ghisellini}, {Giommi}, {Tagliaferri}, {Celotti}, {Chiaberge}, {Fossati}, {Maraschi}, {Tavecchio}, {Treves}, \& {Wolter}}]{costamante01}
{Costamante}, L., {Ghisellini}, G., {Giommi}, P., {et~al.} 2001, \aap, 371, 512

\bibitem[{{Ehlert} {et~al.}(2023){Ehlert}, {Liodakis}, {Middei}, {Marscher}, {Tavecchio}, {Agudo}, {Kouch}, {Lindfors}, {Nilsson}, {Myserlis}, {Gurwell}, {Rao}, {Aceituno}, {Bonnoli}, {Casanova}, {Ag{\'\i}s-Gonz{\'a}lez}, {Escudero}, {Husillos}, {Otero Santos}, {Sota}, {Angelakis}, {Kraus}, {Keating}, {Antonelli}, {Bachetti}, {Baldini}, {Baumgartner}, {Bellazzini}, {Bianchi}, {Bongiorno}, {Bonino}, {Brez}, {Bucciantini}, {Capitanio}, {Castellano}, {Cavazzuti}, {Chen}, {Ciprini}, {Costa}, {De Rosa}, {Del Monte}, {Di Gesu}, {Di Lalla}, {Di Marco}, {Donnarumma}, {Doroshenko}, {Dov{\v{c}}iak}, {Enoto}, {Evangelista}, {Fabiani}, {Ferrazzoli}, {Garcia}, {Gunji}, {Hayashida}, {Heyl}, {Iwakiri}, {Jorstad}, {Kaaret}, {Karas}, {Kislat}, {Kitaguchi}, {Kolodziejczak}, {Krawczynski}, {La Monaca}, {Latronico}, {Maldera}, {Manfreda}, {Marin}, {Marinucci}, {Marshall}, {Massaro}, {Matt}, {Mitsuishi}, {Mizuno}, {Muleri}, {Negro}, {Ng}, {O'Dell}, {Omodei}, {Oppedisano}, {Papitto}, {Pavlov}, {Peirson}, {Perri}, {Pesce-Rollins},
  {Petrucci}, {Pilia}, {Possenti}, {Poutanen}, {Puccetti}, {Ramsey}, {Rankin}, {Ratheesh}, {Roberts}, {Romani}, {Sgr{\'o}}, {Slane}, {Soffitta}, {Spandre}, {Swartz}, {Tamagawa}, {Taverna}, {Tawara}, {Tennant}, {Thomas}, {Tombesi}, {Trois}, {Tsygankov}, {Turolla}, {Vink}, {Weisskopf}, {Wu}, {Xie}, \& {Zane}}]{ixpe23}
{Ehlert}, S.~R., {Liodakis}, I., {Middei}, R., {et~al.} 2023, \apj, 959, 61

\bibitem[{{Eilek}(1979)}]{eilek79}
{Eilek}, J.~A. 1979, \apj, 230, 373

\bibitem[{{Essey} \& {Kusenko}(2010)}]{essey10}
{Essey}, W. \& {Kusenko}, A. 2010, Astroparticle Physics, 33, 81

\bibitem[{{Foreman-Mackey} {et~al.}(2013){Foreman-Mackey}, {Hogg}, {Lang}, \& {Goodman}}]{emcee}
{Foreman-Mackey}, D., {Hogg}, D.~W., {Lang}, D., \& {Goodman}, J. 2013, PASP, 125, 306

\bibitem[{{Galanti} {et~al.}(2020){Galanti}, {Tavecchio}, \& {Landoni}}]{galanti20}
{Galanti}, G., {Tavecchio}, F., \& {Landoni}, M. 2020, \mnras, 491, 5268

\bibitem[{{Ghisellini} {et~al.}(1998){Ghisellini}, {Celotti}, {Fossati}, {Maraschi}, \& {Comastri}}]{ghisellini98}
{Ghisellini}, G., {Celotti}, A., {Fossati}, G., {Maraschi}, L., \& {Comastri}, A. 1998, \mnras, 301, 451

\bibitem[{{Ghisellini} {et~al.}(2017){Ghisellini}, {Righi}, {Costamante}, \& {Tavecchio}}]{ghisellini17}
{Ghisellini}, G., {Righi}, C., {Costamante}, L., \& {Tavecchio}, F. 2017, \mnras, 469, 255

\bibitem[{{Ghisellini} {et~al.}(2010){Ghisellini}, {Tavecchio}, {Foschini}, {Ghirlanda}, {Maraschi}, \& {Celotti}}]{ghisellini10}
{Ghisellini}, G., {Tavecchio}, F., {Foschini}, L., {et~al.} 2010, \mnras, 402, 497

\bibitem[{{Goodman} \& {Weare}(2010)}]{goodman10}
{Goodman}, J. \& {Weare}, J. 2010, Communications in Applied Mathematics and Computational Science, 5, 65

\bibitem[{{Gourgouliatos} \& {Komissarov}(2018)}]{gourgouliatos18}
{Gourgouliatos}, K.~N. \& {Komissarov}, S.~S. 2018, Nature Astronomy, 2, 167

\bibitem[{{Hogg} {et~al.}(2010){Hogg}, {Bovy}, \& {Lang}}]{hogg10}
{Hogg}, D.~W., {Bovy}, J., \& {Lang}, D. 2010, arXiv e-prints, arXiv:1008.4686

\bibitem[{{Kakuwa}(2016)}]{kakuwa16}
{Kakuwa}, J. 2016, \apj, 816, 24

\bibitem[{{Kundu} {et~al.}(2021){Kundu}, {Vaidya}, \& {Mignone}}]{kundu21}
{Kundu}, S., {Vaidya}, B., \& {Mignone}, A. 2021, \apj, 921, 74

\bibitem[{{Larsen} {et~al.}(1985){Larsen}, {Levermore}, {Pomraning}, \& {Sanderson}}]{larsen85}
{Larsen}, E.~W., {Levermore}, C.~D., {Pomraning}, G.~C., \& {Sanderson}, J.~G. 1985, Journal of Computational Physics, 61, 359

\bibitem[{{Lefa} {et~al.}(2011){Lefa}, {Rieger}, \& {Aharonian}}]{lefa11}
{Lefa}, E., {Rieger}, F.~M., \& {Aharonian}, F. 2011, \apj, 740, 64

\bibitem[{{Marscher} \& {Jorstad}(2022)}]{marscher22}
{Marscher}, A.~P. \& {Jorstad}, S.~G. 2022, Universe, 8, 644

\bibitem[{{Matsumoto} {et~al.}(2021){Matsumoto}, {Komissarov}, \& {Gourgouliatos}}]{matsumoto21}
{Matsumoto}, J., {Komissarov}, S.~S., \& {Gourgouliatos}, K.~N. 2021, \mnras, 503, 4918

\bibitem[{{Miller} \& {Roberts}(1995)}]{miller95}
{Miller}, J.~A. \& {Roberts}, D.~A. 1995, \apj, 452, 912

\bibitem[{{Mizuno} {et~al.}(2014){Mizuno}, {Pohl}, {Niemiec}, {Zhang}, {Nishikawa}, \& {Hardee}}]{mizuno14}
{Mizuno}, Y., {Pohl}, M., {Niemiec}, J., {et~al.} 2014, \mnras, 439, 3490

\bibitem[{{Mukherjee} {et~al.}(2021){Mukherjee}, {Bodo}, {Rossi}, {Mignone}, \& {Vaidya}}]{mukherjee21}
{Mukherjee}, D., {Bodo}, G., {Rossi}, P., {Mignone}, A., \& {Vaidya}, B. 2021, \mnras, 505, 2267

\bibitem[{Pareschi \& Russo(2005)}]{pareschi05}
Pareschi, L. \& Russo, G. 2005, Journal of Scientific computing, 25, 129

\bibitem[{{Park} \& {Petrosian}(1996)}]{park96}
{Park}, B.~T. \& {Petrosian}, V. 1996, \apjs, 103, 255

\bibitem[{{Romero} {et~al.}(2017){Romero}, {Boettcher}, {Markoff}, \& {Tavecchio}}]{romero17}
{Romero}, G.~E., {Boettcher}, M., {Markoff}, S., \& {Tavecchio}, F. 2017, \ssr, 207, 5

\bibitem[{{Sciaccaluga} \& {Tavecchio}(2022)}]{sciaccaluga22}
{Sciaccaluga}, A. \& {Tavecchio}, F. 2022, \mnras, 517, 2502

\bibitem[{{Sironi} \& {Spitkovsky}(2011)}]{sironi11}
{Sironi}, L. \& {Spitkovsky}, A. 2011, \apj, 726, 75

\bibitem[{{Stathopoulos} {et~al.}(2023){Stathopoulos}, {Petropoulou}, {Vasilopoulos}, \& {Mastichiadis}}]{stathopoulos23}
{Stathopoulos}, S.~I., {Petropoulou}, M., {Vasilopoulos}, G., \& {Mastichiadis}, A. 2023, arXiv e-prints, arXiv:2308.06174

\bibitem[{{Tavecchio} {et~al.}(2022){Tavecchio}, {Costa}, \& {Sciaccaluga}}]{tavecchio22}
{Tavecchio}, F., {Costa}, A., \& {Sciaccaluga}, A. 2022, MNRAS, in press, arXiv:2207.12766

\bibitem[{{Tavecchio} {et~al.}(1998){Tavecchio}, {Maraschi}, \& {Ghisellini}}]{tavecchio98}
{Tavecchio}, F., {Maraschi}, L., \& {Ghisellini}, G. 1998, \apj, 509, 608

\bibitem[{{Tzavellas} {et~al.}(2023){Tzavellas}, {Vasilopoulos}, {Petropoulou}, {Mastichiadis}, \& {Stathopoulos}}]{tzavellas23}
{Tzavellas}, A., {Vasilopoulos}, G., {Petropoulou}, M., {Mastichiadis}, A., \& {Stathopoulos}, S.~I. 2023, arXiv e-prints, arXiv:2311.06181

\bibitem[{{Urry} \& {Padovani}(1995)}]{urrypadovani95}
{Urry}, C.~M. \& {Padovani}, P. 1995, \pasp, 107, 803

\bibitem[{{Vaidya} {et~al.}(2018){Vaidya}, {Mignone}, {Bodo}, {Rossi}, \& {Massaglia}}]{vaidya18}
{Vaidya}, B., {Mignone}, A., {Bodo}, G., {Rossi}, P., \& {Massaglia}, S. 2018, \apj, 865, 144

\bibitem[{{Zech} \& {Lemoine}(2021)}]{zech21}
{Zech}, A. \& {Lemoine}, M. 2021, \aap, 654, A96

\bibitem[{{Zhou} \& {Matthaeus}(1990)}]{zhou90}
{Zhou}, Y. \& {Matthaeus}, W.~H. 1990, \jgr, 95, 14881

\end{thebibliography}

\appendix
\section{Corner plots}

\begin{figure}[h]
    \centering
    \includegraphics[width = \columnwidth]{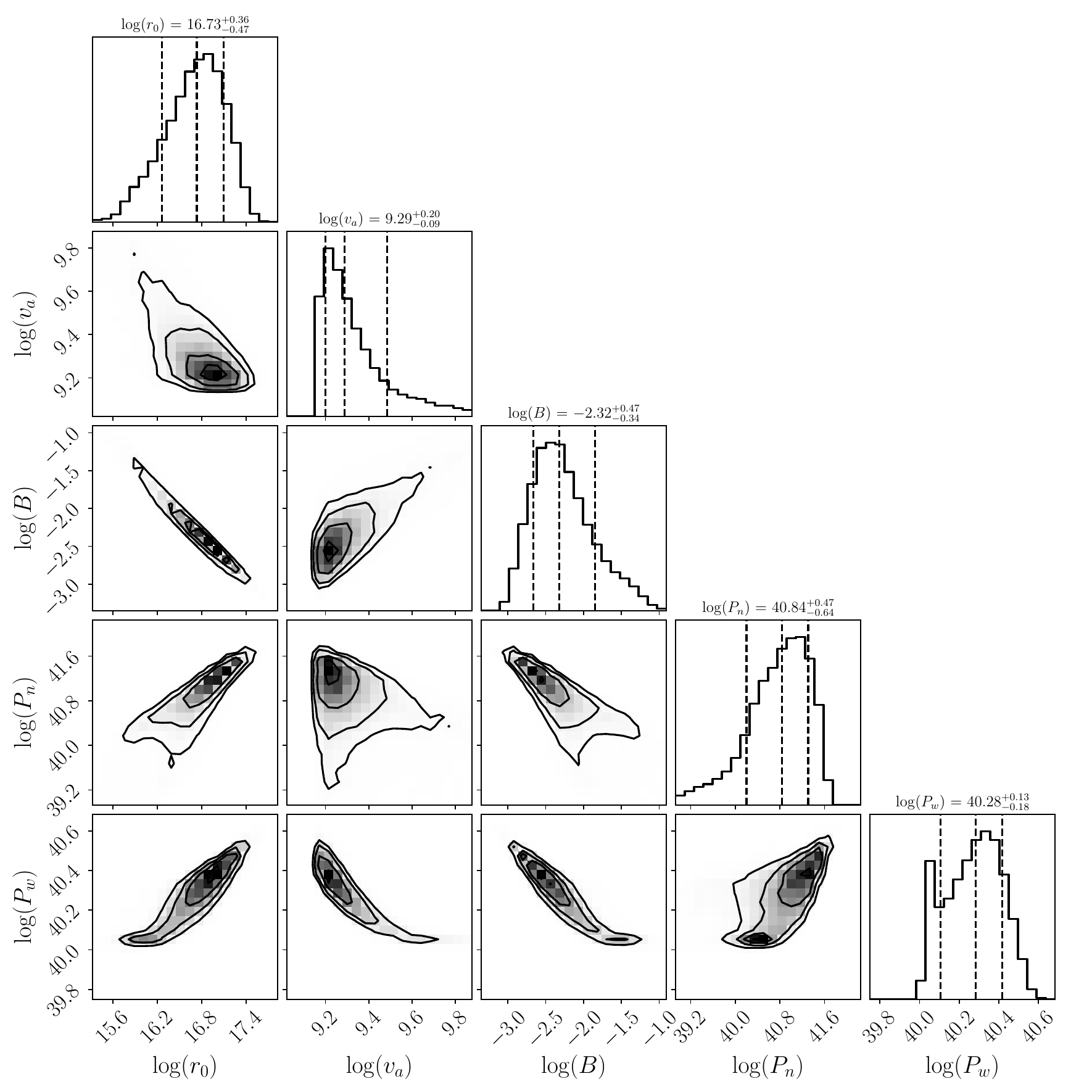}
    \caption{Corner plot of 1ES 0229+200.}
    \label{fig:CP0229}
\end{figure}

\begin{figure}[h]
    \centering
    \includegraphics[width = \columnwidth]{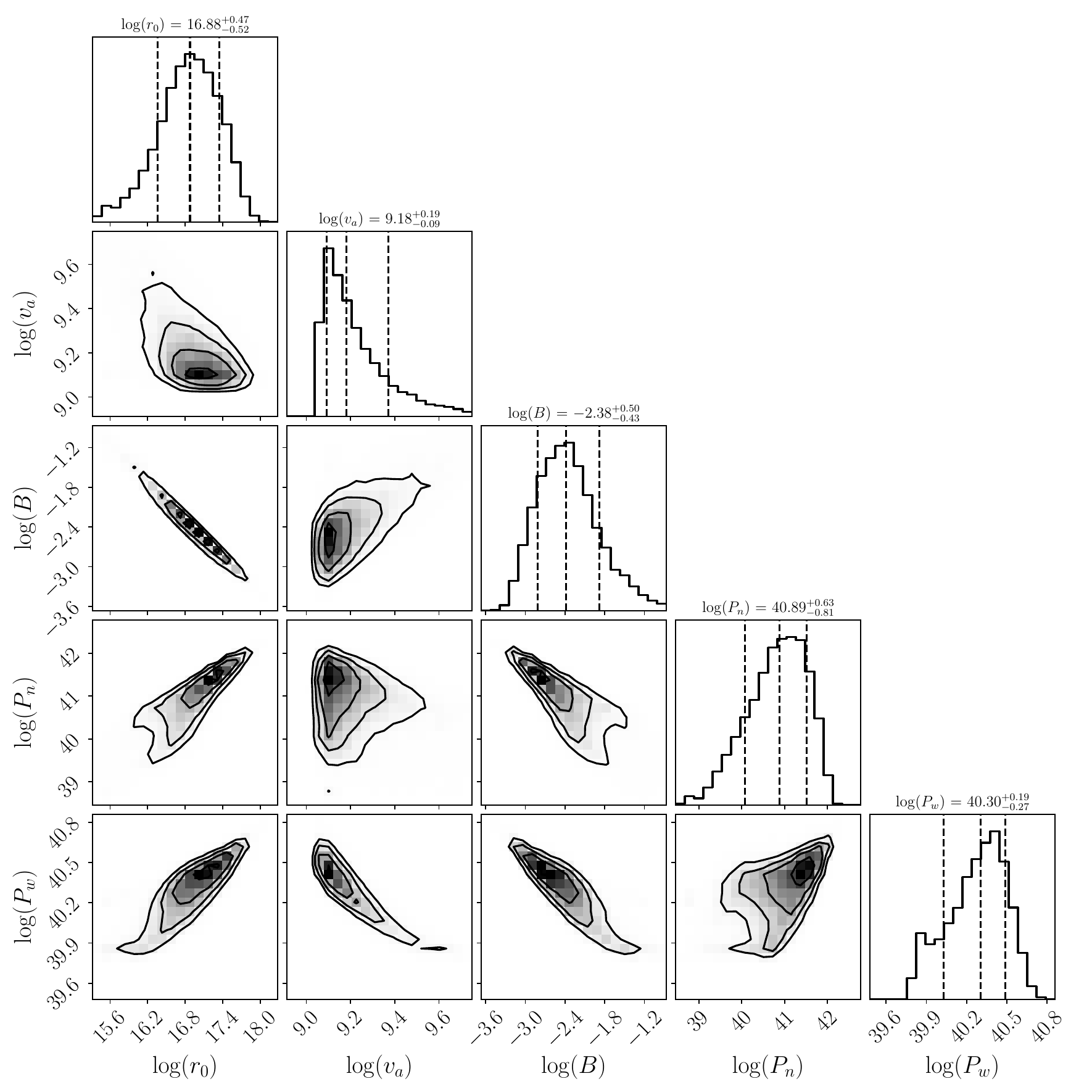}
    \caption{Corner plot of 1ES 0347-121.}
    \label{fig:CP0347}
\end{figure}

\begin{figure}[h]
    \centering
    \includegraphics[width = \columnwidth]{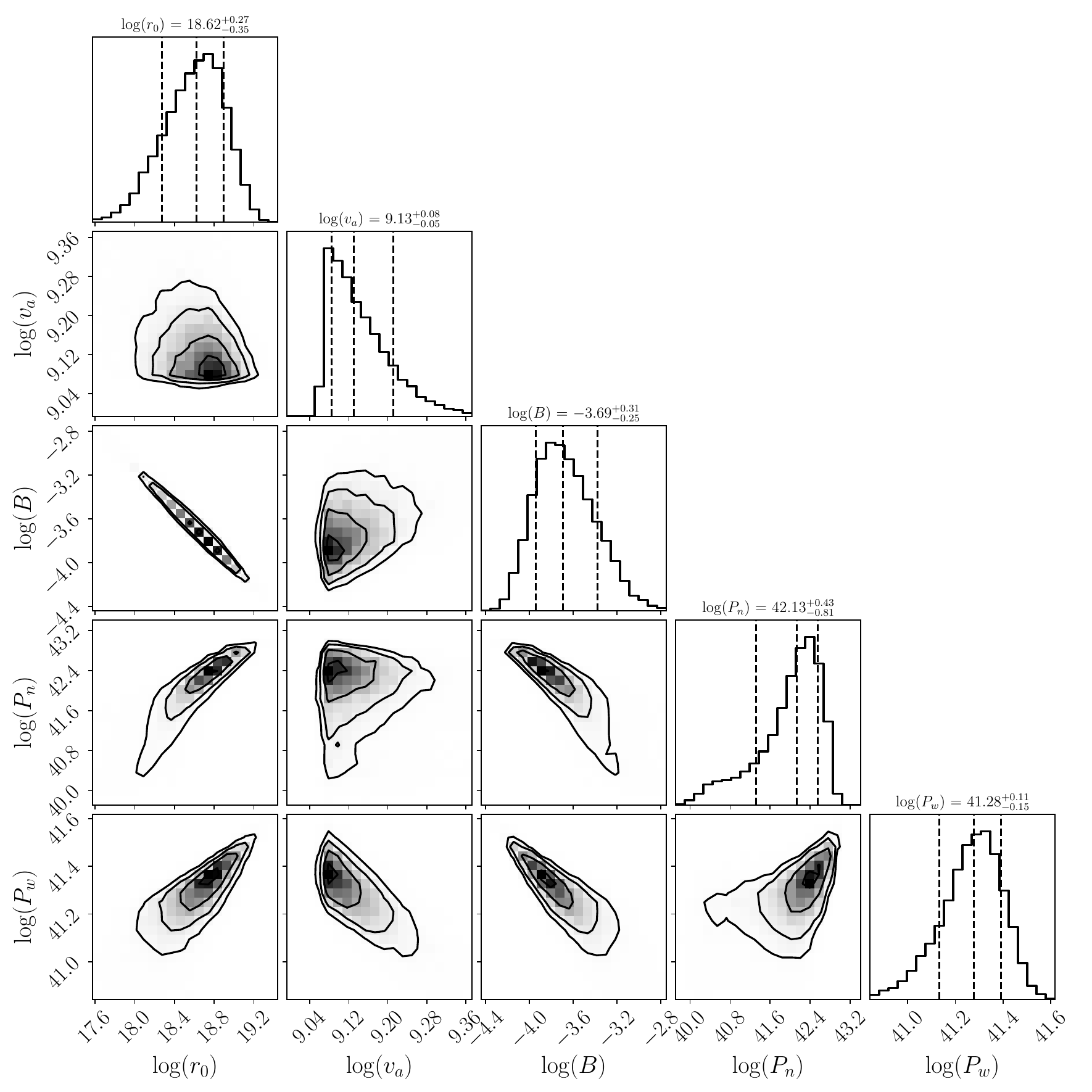}
    \caption{Corner plot of 1ES 1101-232.}
    \label{fig:CP1101}
\end{figure}

\begin{figure}[h]
    \centering
    \includegraphics[width = \columnwidth]{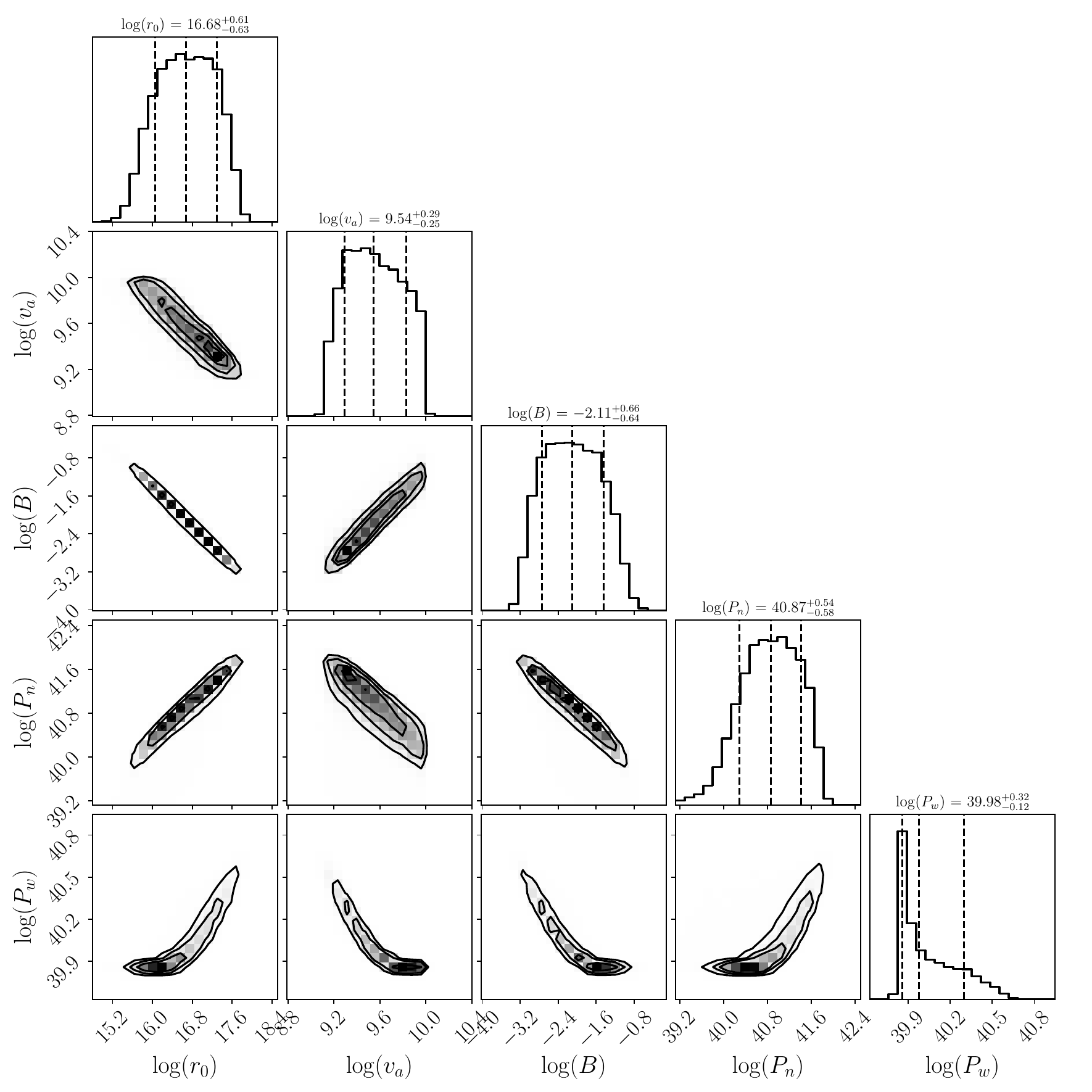}
    \caption{Corner plot of RBG J0710+591.}
    \label{fig:CP0710}
\end{figure}

\clearpage

\section{MCMC of 1ES 1101-232 with $\delta = 40$}

\begin{figure}[h]
    \centering
    \includegraphics[width = \columnwidth]{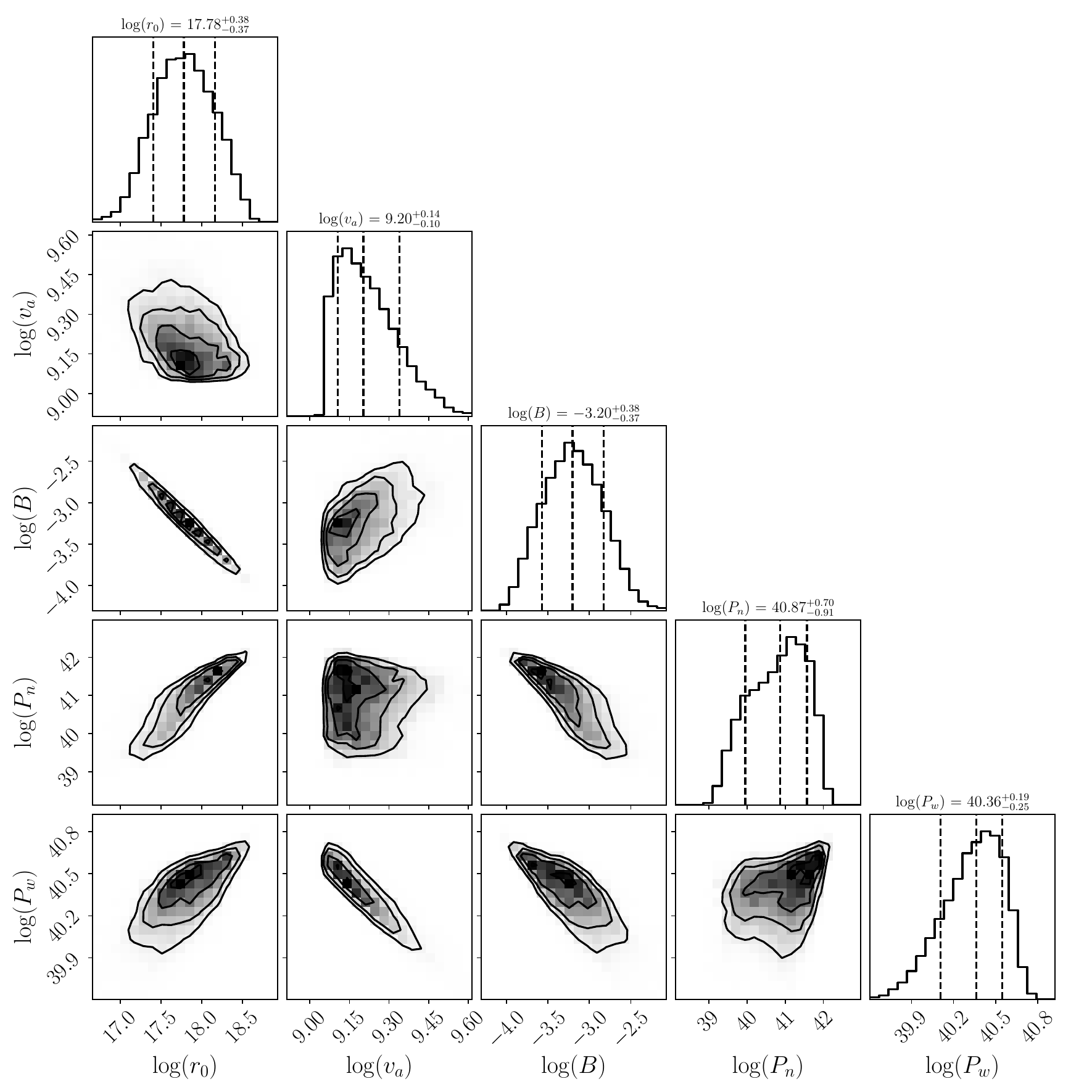}
    \caption{Alternative corner plot of 1ES 1101-232.}
    \label{fig:CP1101}
\end{figure}

\begin{figure}[h]
    \centering
    \includegraphics[width = \columnwidth]{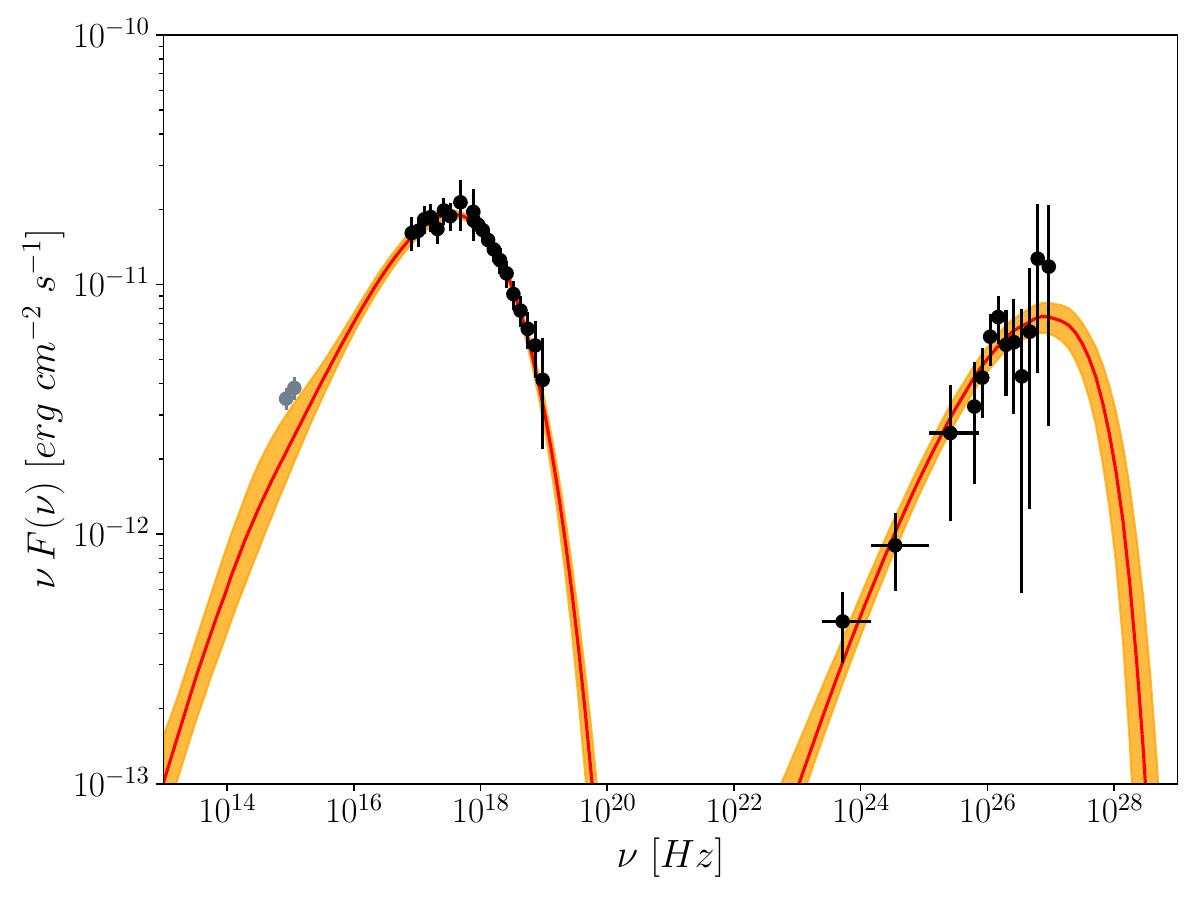}
    \caption{Flux points with their errors (black) of 1ES 1101-232 with $1\sigma$ credible intervals (orange) and the median (red) obtained from the model posterior. Gray points correspond to \textit{Swift}/UVOT data.}
    \label{fig:CP0710}
\end{figure}

\section{Comparison of numerical schemes}

\begin{figure}
    \centering
    \includegraphics[width = \columnwidth]{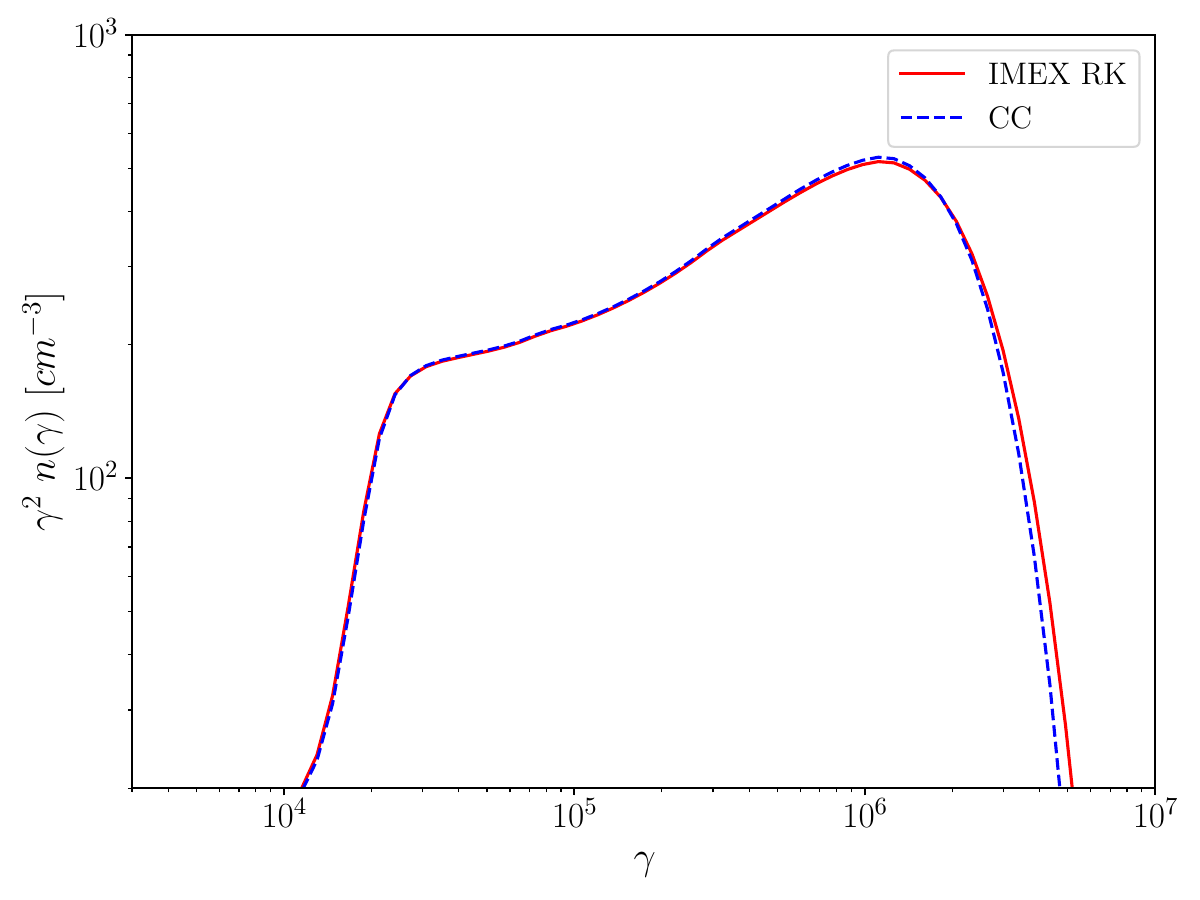}
    \caption{Comparison between two realizations of electron spectrum by IMEX-RK (solid red) and CC (solid blue) using the same parameters, momentum, and wavenumber stepping, but with shorter time stepping for the former.}
    \label{fig:imexrk_vs_cc}
\end{figure}

In previous papers, we used the Chang-Cooper (CC) scheme \citep{chang70, park96} to solve both equations of the system (\ref{eq:system}). We modified the boundary conditions for non-thermal electrons to no flux conditions. This update is negligible for the resulting emission, but it is useful to compare the performances of the CC algorithm with alternative schemes, such as the second-order implicit-explicit Runge-Kutta (IMEX-RK) algorithms \citep{kundu21}. Specifically, we tested the strong stability preserving scheme (2,2,2) of \cite{pareschi05}, but it presents several drawbacks for the scenario we are considering. IMEX-RK can be used exclusively for linear equations, so not for the turbulence equation, making the scheme only first-order accurate. Instead, CC can be adapted for nonlinear equations \citep{larsen85}. Therefore, we are stuck to the same grid accuracy, but, since the advection term is treated explicitly in IMEX-RK, several time steps are necessary, which makes the scheme much slower. However, we used IMEX-RK to test our CC implementation: we used the same momentum and wavenumber grid (i.e., $20$ points per decade) but different time stepping (i.e. ,$100$ time steps for CC and $C = 0.9$ for IMEX-RK, where $C$ is the Courant number). The two schemes produced comparable results (see \fig{imexrk_vs_cc}), confirming the proper functioning of CC implementation; therefore, we decided to keep employing it.

\end{document}